\documentclass[aps,floatfix,showpacs,preprintnumbers,twocolumn]{revtex4}

\usepackage{amssymb}
\usepackage{amsfonts}
\usepackage{amsmath}
\usepackage{graphicx}

\begin{document}

\title{Directed Ratchet Transport in Granular Crystals}
\author{V. Berardi$^{1}$, J. Lydon$^{2,3}$, P.G. Kevrekidis$^{4}$, C. Daraio$^{2,3}$, and
R. Carretero-Gonz\'{a}lez$^{1,}$\footnote{To whom correspondence should be addressed}}
\affiliation{
$^{1}$Nonlinear Dynamical Systems Group\footnote{{\tt URL:} http://nlds.sdsu.edu/},
Computational Science Research Center\footnote{{\tt URL:} http://www.csrc.sdsu.edu/}, and
Department of Mathematics and Statistics,
San Diego State University, San Diego, California 92182-7720, USA
\\
$^{2}$Graduate Aeronautical Laboratories (GALCIT) and Department of Applied Physics,
California Institute of Technology, Pasadena, California 91125, USA
\\
$^{3}$Department of Mechanical and Process Engineering, 
Swiss Federal Institute of Technology (ETH), Z\"urich, Switzerland
\\
$^{4}$Department of Mathematics and Statistics, University of Massachusetts, Amherst,
Massachusetts 01003-4515, USA
}
\date{\today}

\begin{abstract}
Directed-ratchet transport (DRT)  in a one-dimensional lattice of spherical beads, 
which serves as a prototype for granular crystals, is investigated.  We consider a system where the trajectory of the central bead is prescribed by a biharmonic forcing function with broken time-reversal symmetry.  By comparing the mean integrated force of beads equidistant from the forcing bead, two distinct types of directed transport can be observed ---\emph{spatial} and \emph{temporal} DRT.  Based on the value of  the frequency of the forcing function relative to the cutoff frequency, the system can be categorized by the presence and magnitude of each type of  DRT.  Furthermore, we investigate and quantify
how varying additional parameters such as the
biharmonic weight affects DRT velocity and magnitude.  
Finally,  friction is introduced into the system and is found to
significantly inhibit spatial DRT.  In fact,
for sufficiently low forcing frequencies, the friction  may even
induce a switching of the DRT direction.

\end{abstract}

\maketitle

\section{Introduction}

Granular media are large conglomerations of discrete, solid particles, such as sand, gravel, or powder, with unusual, interesting dynamics~\cite{Trujillo2011, Jaeger1996, Luding1995}.  A one-dimensional system of spherical beads in a lattice is one of the simplest representation of granular media substrates, wherein each bead represents a grain of material.   In this approximation, the position of a particular bead is based on forces resulting from its interaction with its two nearest neighbors~\cite{Nesterenko2001}. 
This context has proved especially fruitful for investigating
numerous aspects of the nonlinear dynamic response of such bead chain
systems~\cite{Nesterenko2001,sen08,PGK11}. 
A particular focal
point of emphasis has been on the study of one-dimensional granular
crystals. The availability of a wide variety of materials and bead sizes, as
well as the tunability of the response within the weakly or strongly
nonlinear regime renders such crystals an ideal playground for
the investigation of a variety of fundamental concepts
ranging from nonlinear waves and discrete breathers to shock waves, defect
modes and bifurcation phenomena among many others. However, this
tunability also makes these crystals promising candidates for a wide
variety of engineering applications such as shock and energy absorbing
materials~\cite{dar06,hong05,fernando,doney06}, actuating and focusing
devices~\cite{dev08, spadoni10}, and sound scramblers or
filters~\cite{dar05,dar05b,chiara_nature}.

One aspect that has not been studied, to the best of our knowledge,
in such prototypical granular lattices is that of directed transport
via the so-called ratcheting effect. Directed ratchet transport (DRT), is
defined as the directed transmission of an entity despite the lack of a net
external force acting upon it~\cite{Hanggi09,Rietmann2011}.  
This phenomenon has been associated with applications in
dc current in semiconductors~\cite{Carlin1965}, the
motion of fluxons in Josephson junctions~\cite{Falo99,Falo02,Ustinov2004},
Bose-Einstein condensates~\cite{Rietmann2011,Poletti2008}, cold atoms in
optical lattices~\cite{renzoni1,renzoni2}, among many others.  
{Furthermore, DRT occurring in granular systems has been associated 
with the study of molecular motors~\cite{Julicher1997}}.   
As detailed in
Ref.~\cite{Flach2000}, the emergence of DRT behavior is associated with the
breaking of symmetries, which can be achieved by either a reshaping of the
system's potential or by introducing an external
forcing~\cite{Reimann2002}.  
For instance, DRT is present when a granular material 
is placed in a vertically vibrating sawtooth surface 
profile~\cite{Rapaport:02}.
Typically, DRT is studied (for a single particle
or a collection of particles) when the external input acts on the 
system as a whole~\cite{Falo99,Falo02}. 
Our aim in this work, on the other hand, is to force a {\em single} 
particle to achieve global DRT in the context of granular crystals.

In what follows in section II, we present the basic (Fermi-Pasta-Ulam
type) model that is widely accepted as representing the one-dimensional
dynamics of a granular crystal~\cite{Nesterenko2001,sen08,PGK11} with
parameters that are adapted from recent experiments on the
field such as Refs.~\cite{Boechler2010,Boechler2012}.
We then proceed to use the tunability of the system through
actuating one bead within the chain by means of suitable
biharmonic forcing that will be the source of our DRT through
its induced breaking of time-reversal and half-period time shift
symmetries. In particular, in section III, we will
propose a biharmonic forcing of
the system involving the simplest pair of two frequencies
$(p \omega, q \omega)$ relevant for such DRT (i.e., with
$p, q$ coprimes and $p+q$ odd), namely $p=1$ and $q=2$~\cite{Quintero2010}.
In section IV, we will develop diagnostic quantities evaluating
the relative magnitude of the clearly discernible in our
numerical computations DRT. We will analyze the dependence
of the induced asymmetry in the crystal response on both
the frequency of the drive $\omega$, as well as on the
relative strength of the two terms in the biharmonic forcing,
as controlled by the corresponding parameter $\eta$. The former
analysis will separate different regimes in our observation of
DRT, namely the non-permanent deformation of the crystal
that we will refer to as temporal ratcheting and the permanent
deformation thereof that we will refer to as spatial ratcheting.
The clear distinction between these two regimes is an
especially intriguing feature
of our current setup. The latter analysis (over $\eta$) will
provide a means for optimizing the ensuing transport which can
both be theoretically understood and, in principle, experimentally
exploited. Finally, we consider in section V, the modification
of the above features in the more experimentally realistic setup
incorporating dissipation. We find there that the relevant phenomenology
is modified dramatically, including even a potential reversal of
the direction of the current (for sufficiently low driving frequencies).
Finally, in section VI we summarize our findings and present a number
of directions for future study.

\section{Model and Setup}

\begin{figure}[t]
\centering
\includegraphics[width = 8.5cm]{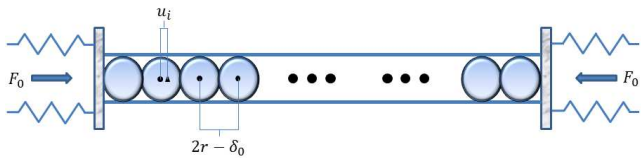}
\vspace{-0.2cm}
\caption{The bead-lattice system with precompression force $F_0$ delivered by springs connected to walls on both ends of the lattice.  The small circular dots represent the location of the center of each bead at its equilibrium position.   Notice that, because of the initial displacement due to precompression, the equilibrium distance between two beads is $2r-\delta_0$.  The triangle represents the position of the center of the $i$th bead after displacement $u_i$. }
\label{fig:BeadLattice}
\end{figure}

To ensure that the beads remain in contact, we consider a horizontal lattice that is precompressed on both ends with a force $F_0$ resulting in a static bead displacement $\delta_0$ (see Fig.~\ref{fig:BeadLattice}).
The existence of the precompression also serves to ensure that a linear
spectrum of excitations exists in the lattice (see details below).
With these considerations, based on the Hertzian law of spherical point
contacts, a system comprised of $N$ identical beads can be described by the following Newtonian equation~\cite{Nesterenko2001}:
\begin{equation}\label{EOMComp}
m\ddot{u}_i = A[\delta_0+u_{i-1}-u_i]_+^\frac{3}{2} - A[\delta_0+u_{i}-u_{i+1}]_+^\frac{3}{2}
\end{equation}
where $[Y]_+ = {\rm max}\left\{0,Y\right\}$, $m$ is the bead mass, $u_i$ is the displacement of the center of the $i$th bead from its equilibrium position, and $A$ is the Hertzian constant calculated as
\begin{equation}
A\equiv\frac{2E\sqrt{r}}{3\sqrt{2}(1-\nu^2)},
\end{equation}
where $r$, $E$, $\nu$ are, respectively, the bead's radius, Young's elastic modulus, and Poisson's ratio.  The static displacement is $\delta_0 = \left(\frac{F_0}{A}\right)^{\frac{2}{3}}$.  In line with the experiments of Refs.~\cite{Boechler2010,Boechler2012}, the parameter values listed in Table~\ref{tab:DefParam} were used.
\begin{table}[hbt]
\centering
\caption{Default Parameters for Bead-Lattice System.}\label{tab:DefParam}
\begin{tabular}{|c|c|c|}
\hline
{Parameter}          & {Symbol} & {Default Value}\\ \hline \hline
Mass                 &  $m$     & 28.84 g\\ \hline
Radius               &  $r$     & 9.53 mm\\ \hline
Poisson's ratio      &  $\nu$   & 0.3\\ \hline
Young's modulus      &  $E$     & 0.193 $\frac{g}{{\rm mm}\,\mu{\rm s}}$\\ \hline
Precompression force &  $F_0$   &  5 N\\ \hline
\end{tabular}
\end{table}


{As we will show later, a} 
critical factor affecting the presence/type of DRT is the 
acoustic phonon band
cutoff frequency.
Plane wave solutions to the system follow the dispersion relation 
$\bar{\nu}(\alpha)^2 = {\frac{3}{2}A\sqrt{\delta_0}\sin^2(\pi\delta_0 \alpha)}
/({m\pi^2})$~\cite{Brillouin1946}, where $\alpha$ is the wave number and 
$\bar{\nu}$ is the temporal frequency.
We see that this relationship is periodic (with period ${1}/{\delta_0}$) and that there is a cutoff frequency $\bar{\nu}_c$, above which plane wave solutions cannot propagate.  The maximal frequency value occurs at the boundaries of the $-\frac{1}{2\delta_0}\leq \alpha\leq\frac{1}{2\delta_0}$ interval, which correspond to the smallest allowable wavelength.  Substituting $\alpha=\pm\frac{1}{2\delta_0}$ into the dispersion relation yields the cutoff frequency,
\begin{equation}
\bar{\nu}_c = \frac{1}{\pi}\sqrt{\frac{\frac{3}{2}A\sqrt{\delta_0}}{m}}.
\end{equation}
Therefore, $0<\bar{\nu}<\bar{\nu}_c$ defines the range of propagating frequencies, called the acoustic band.  Frequencies $\bar{\nu}>\bar{\nu}_c$ lie within the band gap and cannot propagate through the lattice as plane waves.   With the parameter values given in Table~\ref{tab:DefParam}, we have $\bar{\nu}_c = 6.42$~kHz.  In terms of angular frequency, $\omega_c=2\pi\bar{\nu}_c=$~40.31~${\mathrm{rad}}/{\mathrm{ms}}$, which is the critical frequency used from this point forward. Notice that all the frequencies that will be mentioned hereafter
will be measured in ${\mathrm{rad}}/{\mathrm{ms}}$.

\section{Biharmonic Forcing}

Typically, DRT behavior is observed in the velocity of a single 
particle or in that of a coherent structure such as a solitary 
wave~\cite{Rietmann2011,Flach2000,Reimann2002,Quintero2010,Jesus10}.  
In the case of the granular crystal
though, DRT will be observed (and examined) throughout the system as a whole.   Consider a lattice where each bead begins at its equilibrium position with no initial velocity.  
To introduce energy into the system, the $i^*$th bead, 
located at the center of the lattice, is controlled by the 
following biharmonic, periodic function
\begin{equation}\label{Biharmonic2}
u_{i^*}(t) = a\left[\eta\sin(\omega(t+\phi))+(1-\eta)\sin(2\omega(t+\phi))\right],
\end{equation}
where $t$ is time, $a$ is the amplitude, $\omega$ is the 
frequency, $\eta$ is the biharmonic weight, and $\phi$ is a phase.  
To maintain uniformity on each side of $i^*$th bead, we assume $N$ is odd.  
The motivation for choosing to control the displacement
of the central bead is that we envisage the possibility of 
doing experimental DRT studies in the future where the
position of the central bead will be controlled by an actuator.
An essential characteristic of this functional form is that it has a 
zero-integral over one period, indicating
that the function is not biased in any direction. 
In other words, the $i^*$th bead's temporal center of mass,
relative its equilibrium position, is zero. 
Consequently, any directed behavior observed must be attributed 
to DRT rather than a preferential direction for the input.
It is relevant to note that the prescription of the motion
of the $i^*$th bead is 
tantamount to introducing
a force,
with the same characteristics, into its nearest neighbors
through the equations of motion (\ref{EOMComp}).

\begin{figure}[htb]
\vspace{-0.2cm}
\centering
\includegraphics[width = 6.5cm]{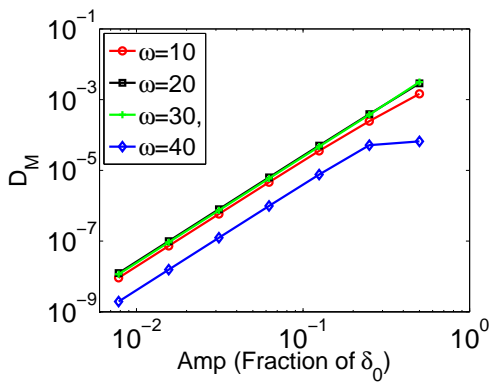}
\caption{DRT magnitude vs.~forcing amplitude $a$.  
The default parameters in Tables~\ref{tab:DefParam} 
are used with 
$\eta={4}/{9}$ and $\phi$ averaged over 16 values.}
\label{fig:VaryAmp}
\end{figure}

For $\eta=\{0,1\}$  each bead orbit on one side of the lattice corresponds
directly to an orbit on the other side of the lattice traveling in the opposite direction, translated by a half-period delay.  These orbits exactly cancel each other out and thus there is no DRT. 
However, when $\eta\in(0,1)$, the symmetry of $u_{i^*}(t)$ is broken 
and DRT can occur  \cite{Quintero2010}.

The system is numerically solved using a fourth-order Runge-Kutta scheme.  The conservation of total energy is used to determine an appropriate time step.  The final integration time $\tau$ varies based on the frequency $\omega$, but is always selected so that it is an integer multiple of  $T$, the period of $u_{i^*}(t)$.  $N$ also varies with $\omega$, but is always sufficiently large so that energy from $i^*$th bead's oscillation never reaches the 1st or $N$th bead.

We consider values of $\omega$ ranging from 10 to 40
and set $\eta$ equal to ${4}/{9}$.  In Section~\ref{sec:IP&R}, 
we demonstrate that these parameters result in DRT towards the 
right-hand side of the lattice. 
It is possible to change this 
direction by adding an additional phase 
mismatch between the two harmonics 
of the driver $u_{i^*}(t)$ (results not shown here). 
Figure~\ref{fig:VaryAmp} illustrates DRT magnitude (quantification is discussed in Section~\ref{sec:IP&R}) for values of $a$ ranging from $\delta_0/128$ to $\delta_0/2$.  
These relatively small values of the forcing amplitude $a$ ensure 
that a (relatively) small amount of energy is introduced into the 
system so the beads always remain in contact with each other.  
The relationship between $a$ and the DRT magnitude is clearly nonlinear, 
as the average slope of the lines in the log-log plot in 
Fig.~\ref{fig:VaryAmp} is about 2.9725, which indicates 
an essentially cubic (gain) relationship.  
Based on these findings, the remaining 
simulations have $a\equiv\delta_0/4$
in order to exploit most of the nonlinear gain but also
avoid losing contact between all beads for all times. 

\section{Force Profiles and Ratcheting}\label{sec:IP&R}

In order to quantify the DRT displayed by the bead chain,
let us define the quantity $I_i(t)$ corresponding to the 
average of the Hertzian forces on either side of the $i$th bead
integrated over time.
This choice of DRT measure is inspired by the fact that
a piezo embedded inside a bead precisely measures the
average of the Hertzian forces felt by the adjacent beads.
It is important to mention at this stage that DRT could
be captured using many possible measures. In fact, we also
used, instead of $I_i(t)$, the actual forces acting on each
bead and other combinations thereof and the results are
qualitatively similar (results not shown here).
We should point out that it is essential to consider 
the entire space of possible phases $\phi$ in $u_{i^*}(t)$.
This allows the full spectrum of the function to be sampled
without biasing any direction based on the initial phase of
the driver.
To do this in our numerical experiments, we consider sixteen values of $\phi$, equally spaced throughout one period of $u_{i^*}(t)$ and define $\bar{I}_i(t)$
as the average of $I_i(t)$ over these phases.  
%

\begin{figure}[t!]
\vspace{-0.2cm}
\includegraphics[width = 4.2cm]{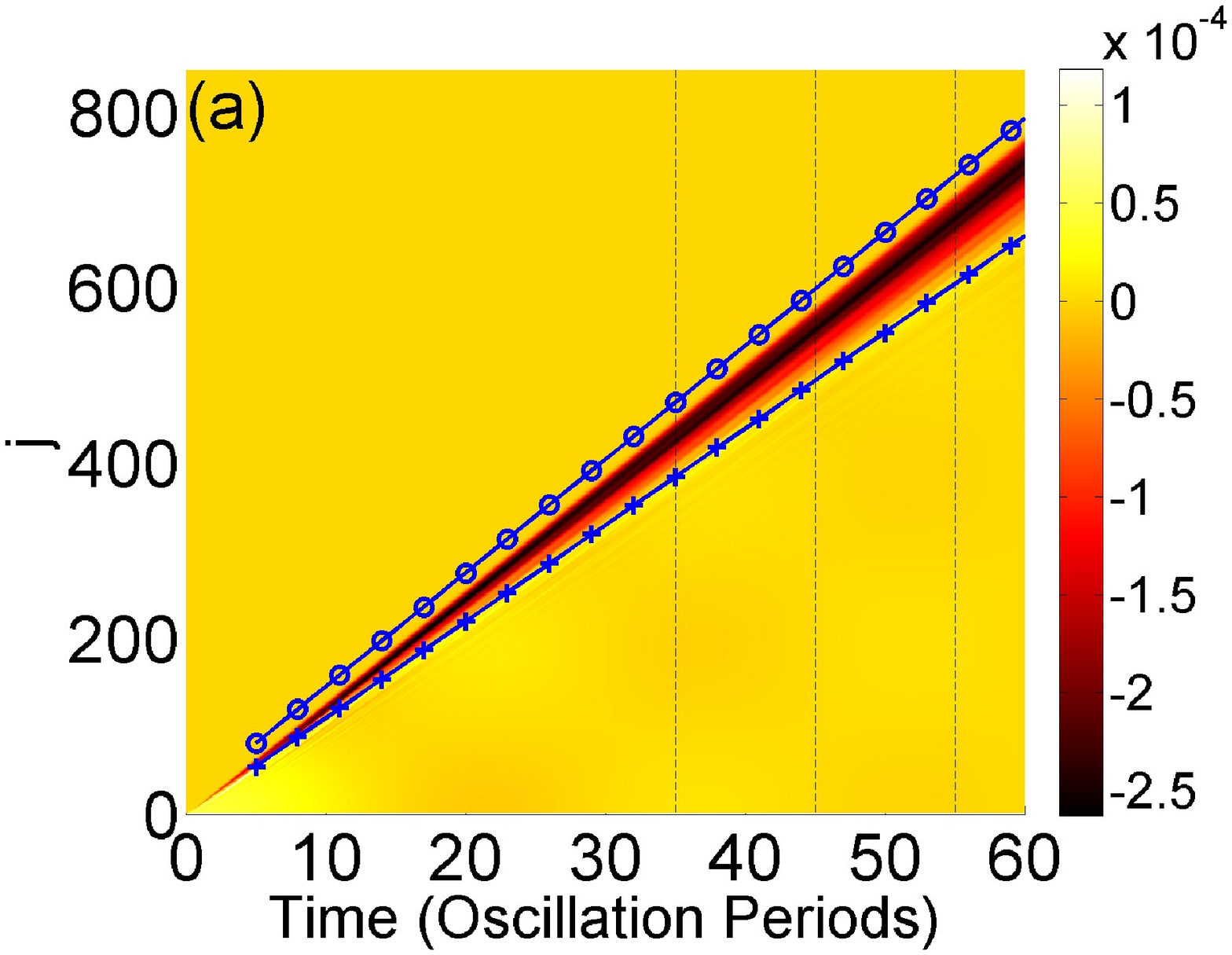}
\includegraphics[width = 4.2cm]{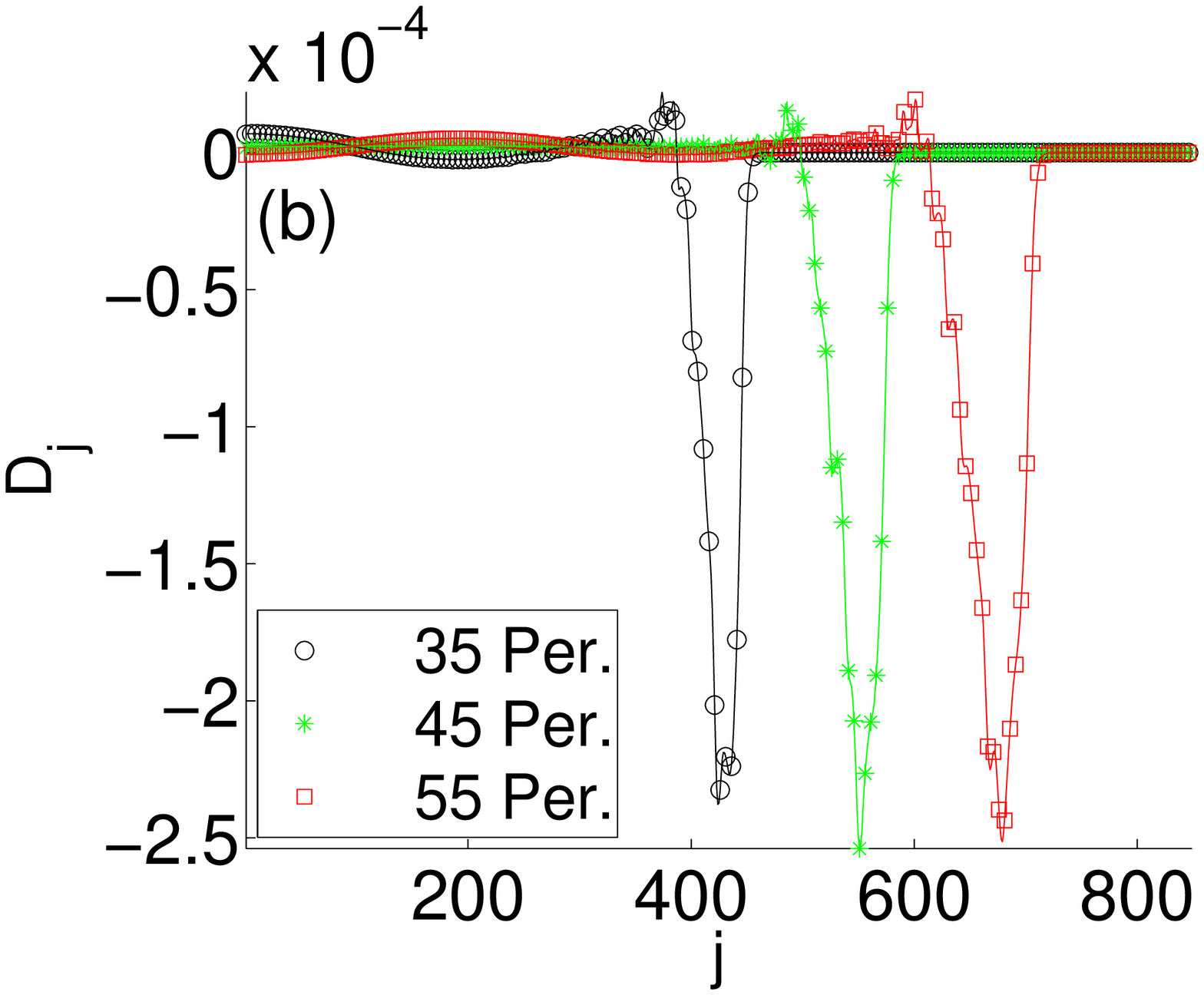}\\
\includegraphics[width = 4.2cm]{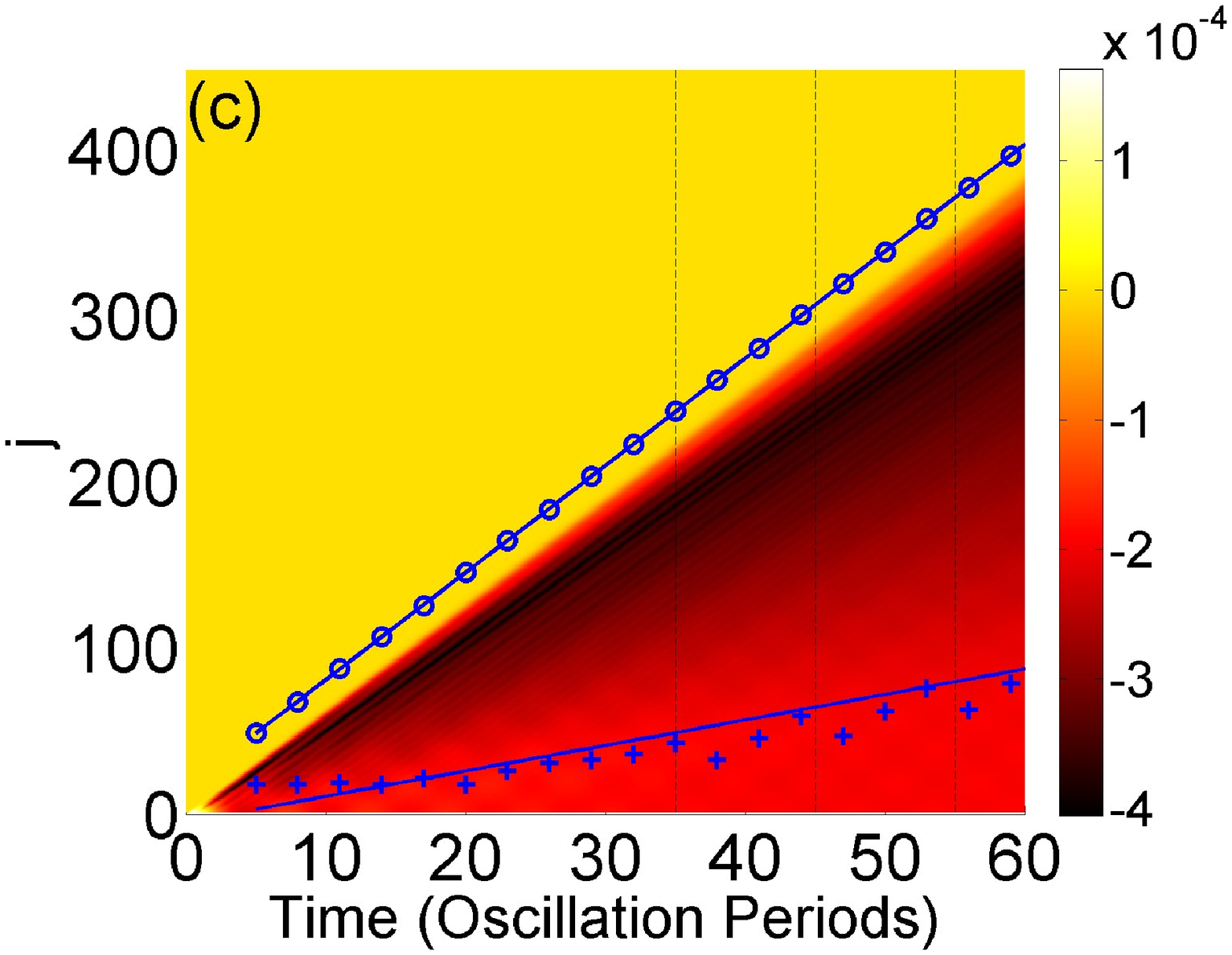}
\includegraphics[width = 4.2cm]{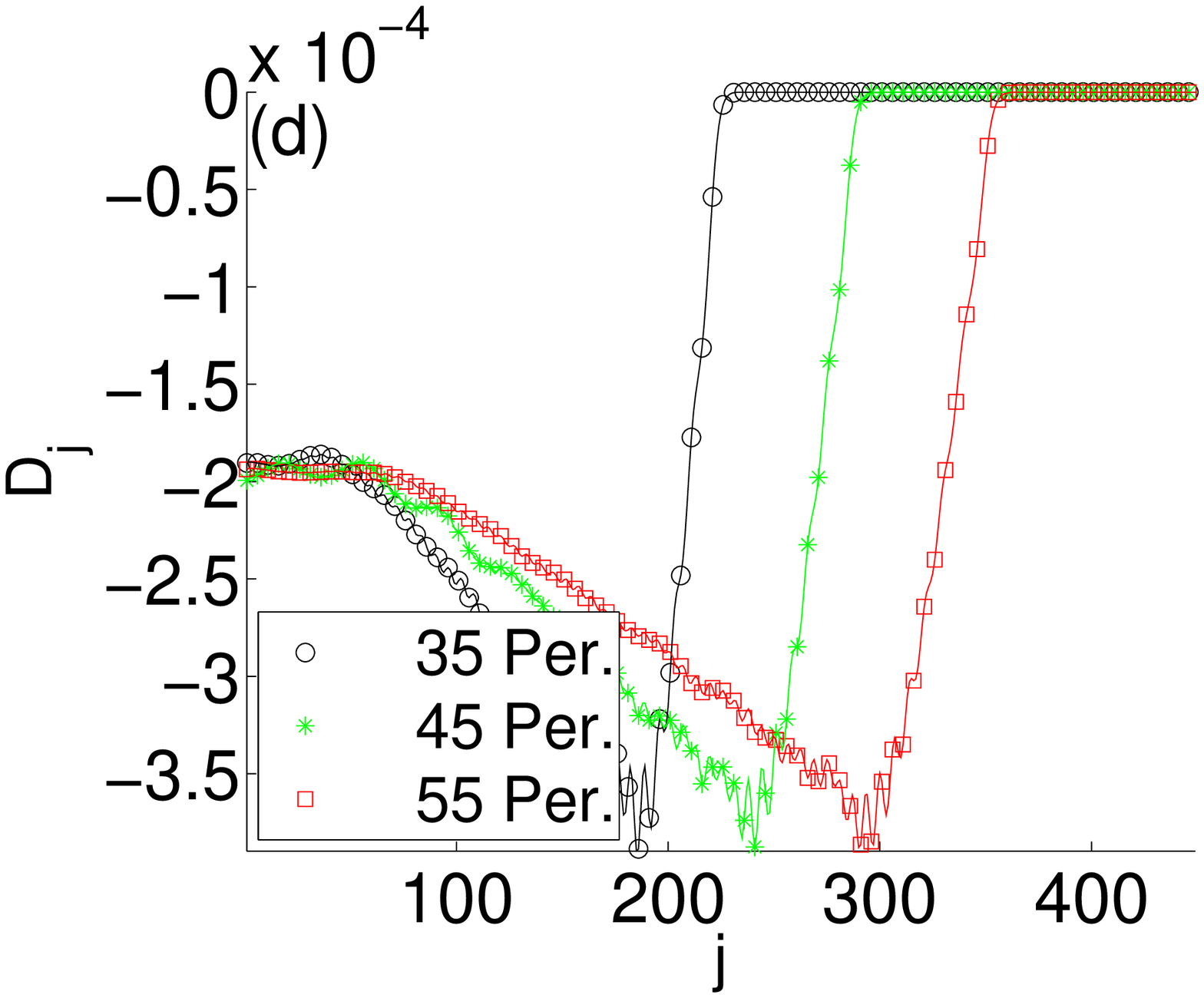}\\
\includegraphics[width = 4.2cm]{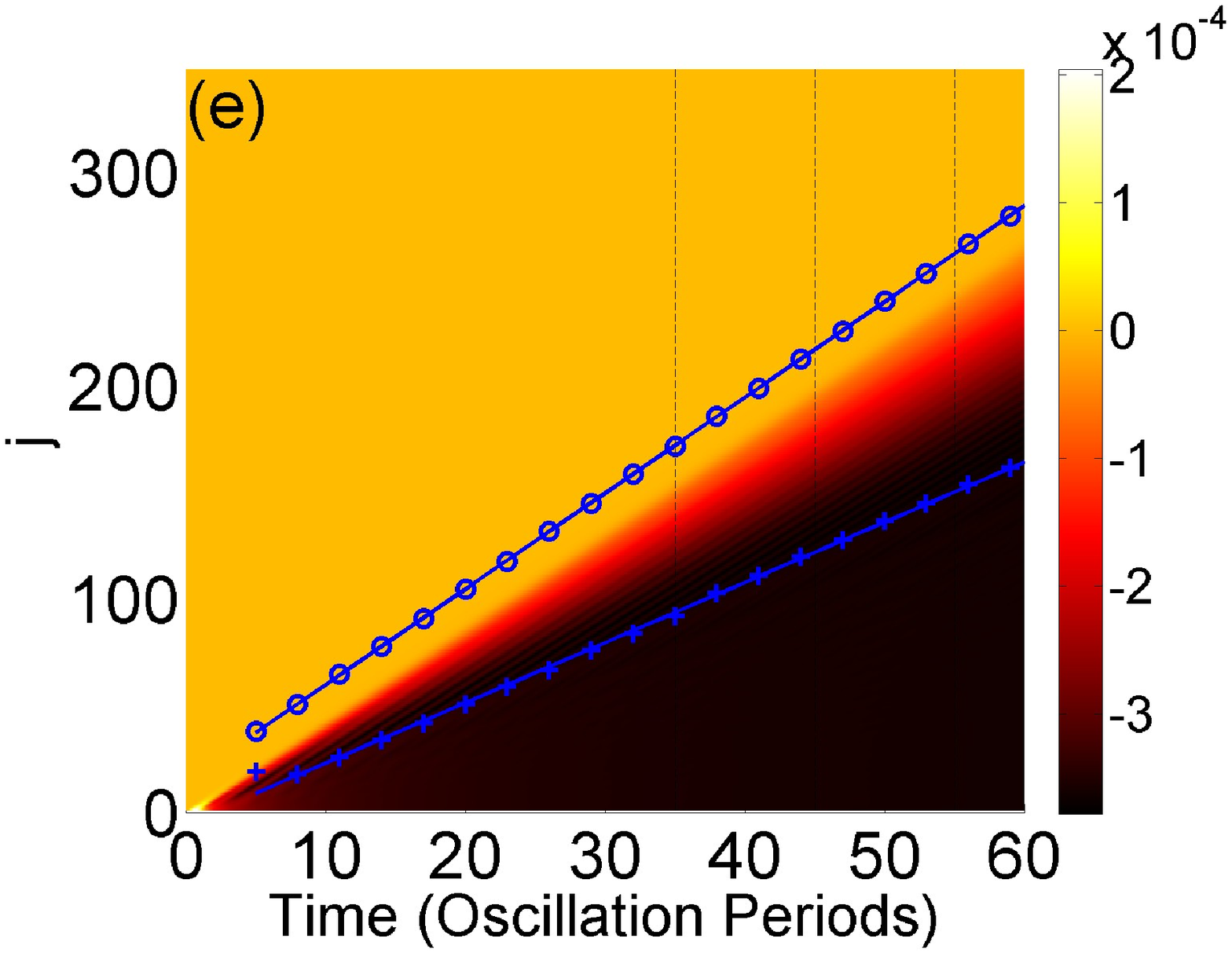}
\includegraphics[width = 4.2cm]{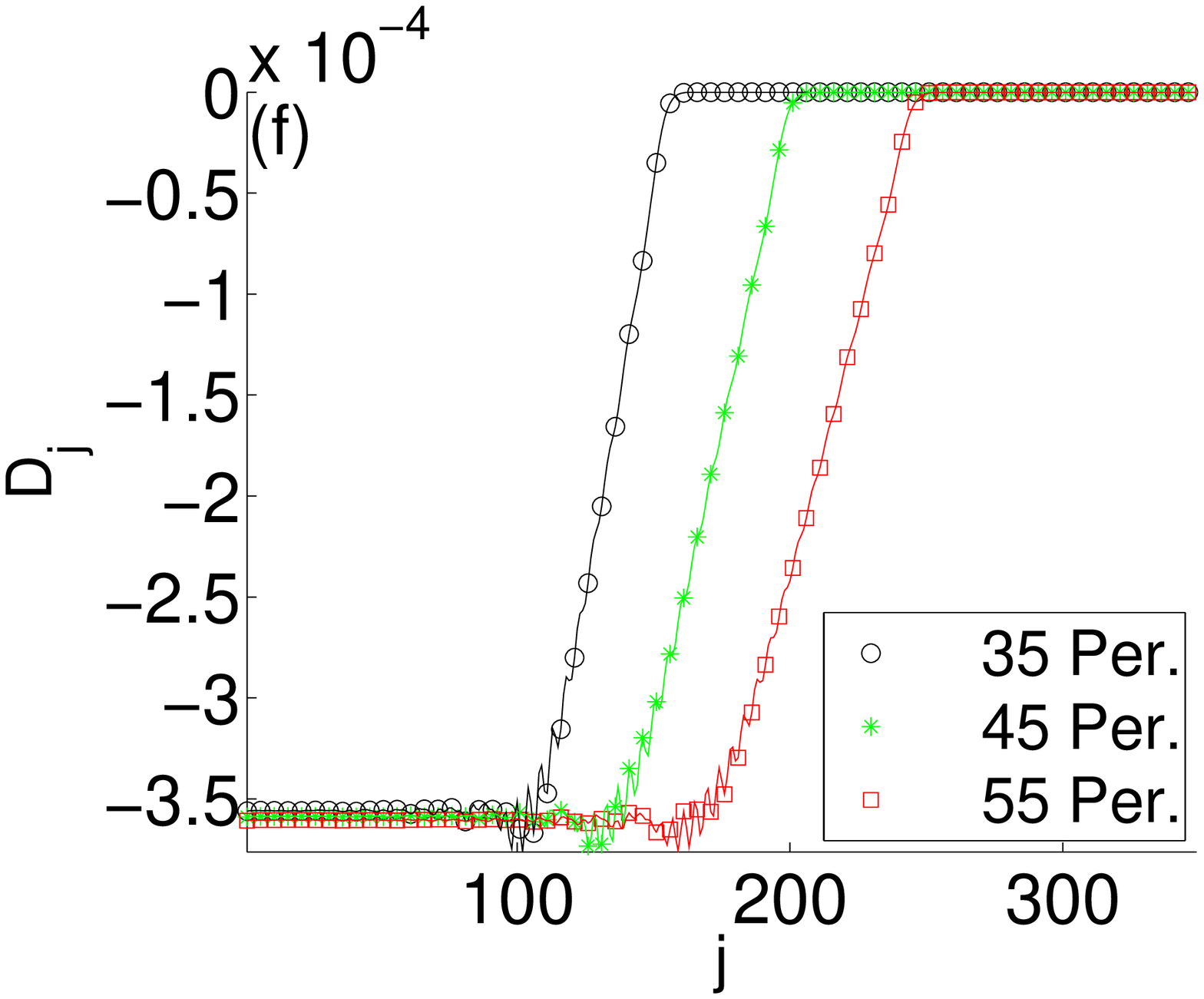}\\
\includegraphics[width = 4.2cm]{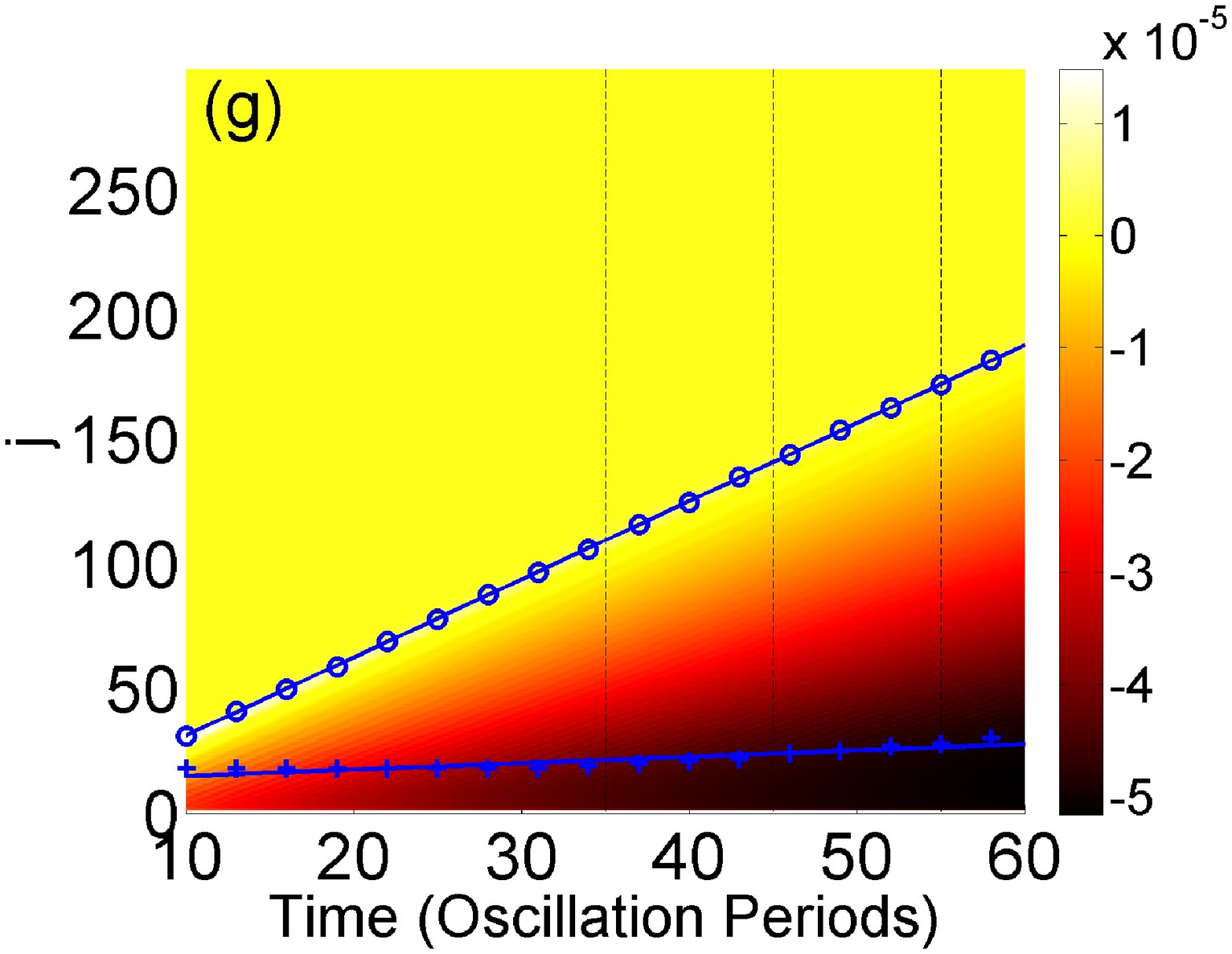}
\includegraphics[width = 4.2cm]{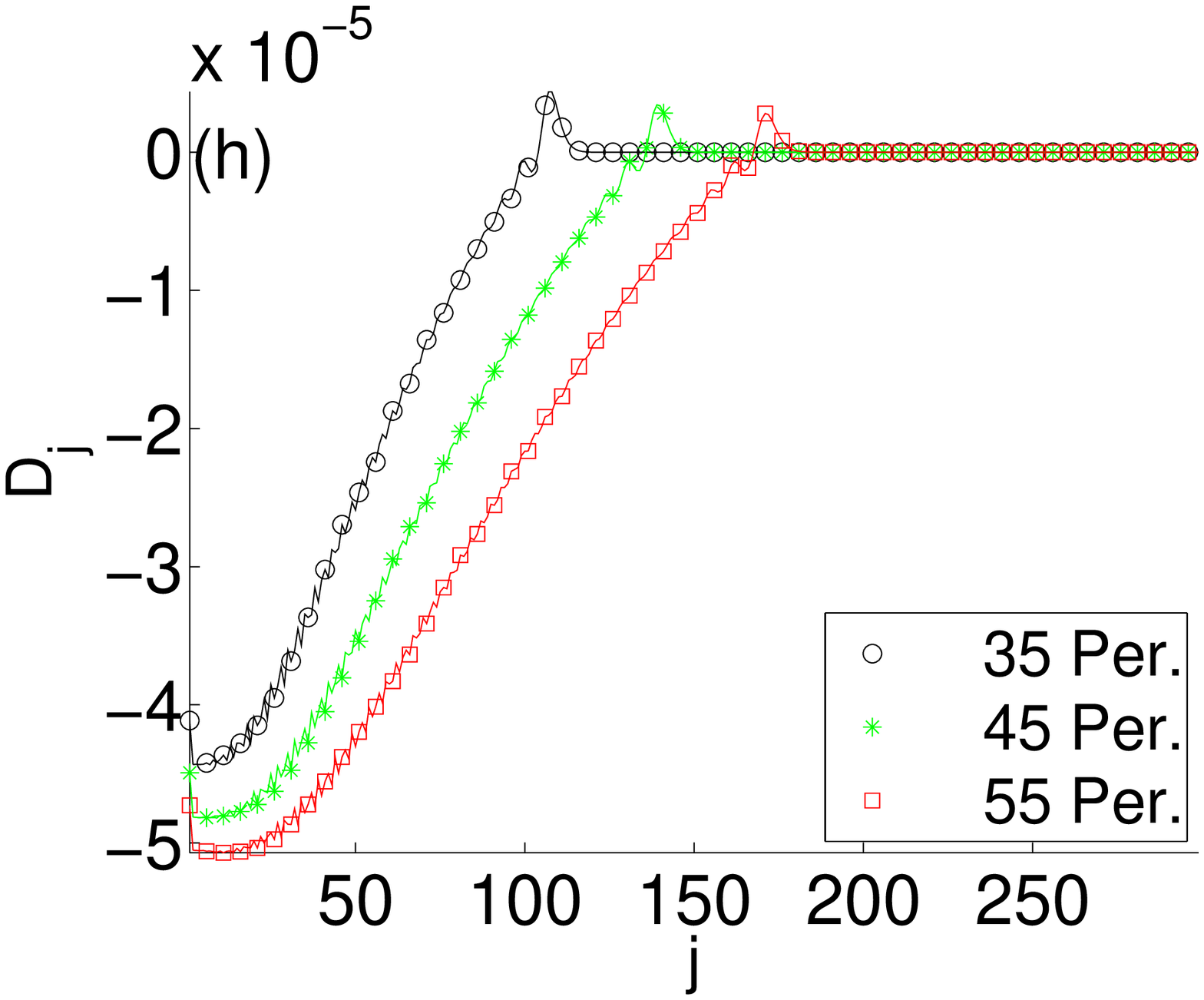}
\caption{Spatio-temporal evolution of the asymmetry indicator
($D_j$, see Eq.~(\ref{ImpDif})) [left panels] and its spatial dependence
for particular times [right panels].
(a),(b): $\omega=10$,  (c),(d): $\omega=20$;  (e),(f): $\omega=30$,  (g),(h): $\omega=40$.  The outer-cone horizon was calculated by identifying the first value for $j$, where the wave had not yet reached and is denoted by circles.  The crosses indicate the inner-cone horizon.  For $\omega = \{10,20,30\}$, the location was determined by identifying the first value of $j$ to the left of the minimum of $D_j$ where the derivative between beads is approximately zero.  For $\omega=40$, the minimum value of $D_j$ identifies the inner-cone horizon.  To improve clarity, only every third period is shown.  A linear least-squares best fit line is depicted for these locations; the slope of the line is the cone velocity.  The vertical dashed lines indicate the particular times for which the
spatial asymmetry profile is illustrated on the right panels.  To improve clarity in the right panels, only every fifth value of $j$ is shown. The same parameters as Fig.~\ref{fig:VaryAmp} were used.}
\label{fig:ImpProf}
\end{figure}

To create profiles which will allow DRT behavior to be observed, we 
compare the normalized difference of $\bar{I}_i(t)$ for pairs of 
beads equidistant from the center bead, that is
\begin{align}\label{ImpDif}
D_j = \frac{\bar{I}_{i^*-j}-\bar{I}_{i^*+j}}{\bar{I}_{i^*+j}},
\end{align}
where $j\in(1,\frac{N+1}{2}$). By monitoring $D_j$ over time, 
a DRT profile is constructed (see Fig.~\ref{fig:ImpProf}).

In Fig.~\ref{fig:ImpProf}, for each value of $\omega = \{10,20,30,40\}$, a
contour plot illustrating the DRT profile as a function of $j$ and $t$ is provided.  Additionally, the right panels show the asymmetry
indicator profile of Eq.~(\ref{ImpDif})) at a set of particular times,
written in
terms of oscillations of the center bead.  A non-zero value of $D_j$
indicates the preferential transport of force in one direction, that is, DRT.
We see that, after transient behavior, all  significant values of $D_j$ are negative, indicating the presence of DRT towards the right hand side of the
lattice (see Eq.~\eqref{ImpDif}).

Figures~\ref{fig:ImpProf}(a),(b), illustrate the DRT profiles
for $\omega=10$, where both $\omega$ and $2\omega$ are below $\omega_c$.  
After a transient time interval, we observe a DRT ``wave'', or cone, 
advancing as time progresses 
(and leaving no DRT behind it).
For this reason we call this behavior \emph{temporal DRT}.  For a given 
time, let $O$ denote the value of $j$ at the outer edge of the cone
(that effectively travels at the speed of sound within the medium), $\cal{I}$ denote the value of $j$ at the inner edge of the cone, and $M$ denote the value of $j$ for which $\vert D_j\vert$, and therefore DRT, is maximal.  For $j>O$, we have $D_j = 0$ since the energy from the forcing function has not yet reached beads offset this far from the center bead. For the region defined by  $M\leq j\leq O$, there exists a positive, approximate-linear relationship
(with respect to $j$)
 describing the magnitude of the DRT.  Similarly,  for $\cal{I}$\,$\leq j\leq M$, there exists a negative approximate linear relationship describing DRT magnitude.  The  DRT wave has already moved through the region defined by $0<j<\cal{I}$.  The characteristic property of this class of behavior, which we call Class I, is that for $0<j<\cal{I}$, $D_j\approx0$, indicating DRT is {\it no longer
present} shortly after the wave has left a region.

Class II behavior is observed by considering a forcing frequency of 20, as shown in  Figs.~\ref{fig:ImpProf}(c),(d).  Here, $\omega_c$ is greater than $\omega$ and slightly larger than $2\omega$.  After allowing for transient time, the regions defined by $j>O$,  $M\leq j\leq O$, and $\cal{I}$\,$\leq j\leq M$, display the same qualitative behavior as the previous case.  However, what
clearly distinguishes this region is that for $j<\cal{I}$ we see a qualitatively different result, namely, for each $j$, $D_j$ is approximately equal to a nonzero constant.  This is indicative of an equilibrium DRT state defined by the spatial extent of the region through which the ratcheting wave has already passed.  This effect and the ``kink''-like pattern that it leads to
(rather than the pulse like structure of Class I) in the
context of the asymmetry indicator $D_j$ is hereafter referred to
as \emph{spatial DRT}.
This fundamental distinction of regimes of temporal and spatial
DRT is, arguably, one of the most interesting traits observed herein, and to our knowledge, has not been reported before, although we believe that it
should be more general than the particular realization considered herein.

In Figs.~\ref{fig:ImpProf}(e),(f), $\omega$ = 30 and thus $\omega<\omega_c$ 
and $2\omega>\omega_c$. As a result, a different behavior that
will be characterized hereafter as belonging to Class IIIA is observed.
The features are similar to Class II, but now the magnitude of the spatial
DRT is approximately equal to $\vert D_M\vert$, the maximal temporal DRT
magnitude.  
Otherwise said, the tail of the kink associated with the
asymmetric deformation of the lattice, rather than having the linear
profile of Class II, it is essentially flat.
%
As shown in Figs.~\ref{fig:ImpProf}(g),(h), where $\omega=40$, as $\omega$ approaches $\omega_c$, the DRT profile remains qualitatively the same, but the slope of the (approximately) linear relationship for $M\leq j\leq O$ decreases.
In our kink-based visualization of the corresponding $D_j$'s, this regime
is associated not with the translation of the structure over the
lattice which roughly
preserves its shape, as in Class IIIA. Instead, it appears associated
predominantly with the
widening of the relevant spatial structure in this regime that we will
refer to as Class IIIB.

\begin{figure}[htb]
\vspace{-0.2cm}
\centering
\includegraphics[width = 6.5cm]{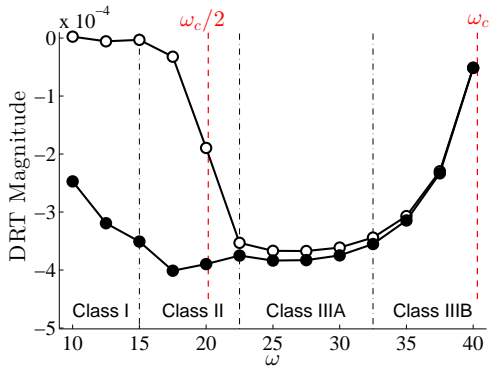}
\caption{The magnitude of spatial and temporal DRT as a function of $\omega$.  The open dots represent the magnitude of spatial DRT (as measured
by $D_1$) while the filled dots are the magnitude of temporal DRT
(as measured by $D_M$; see also the text).
The same parameter values and conditions as Fig.~\ref{fig:Freq-Vel} were used.}
\label{fig:Freq-DRT}
\end{figure}

\begin{figure}[htb]
\vspace{-0.2cm}
\centering
\includegraphics[width = 6.5cm]{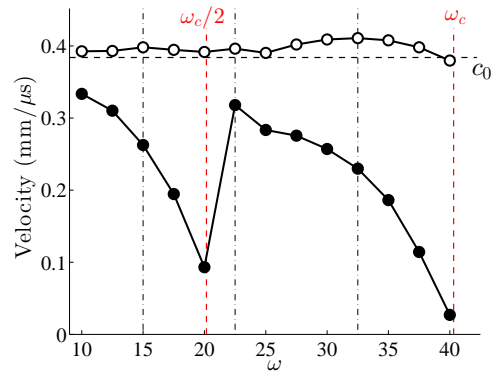}
\caption{Outer- and inner-cone velocities as a function of $\omega$.  The open dots represent the inner-cone velocities while the filled dots are the
outer-cone velocities.  The horizontal dotted line corresponds to the sound velocity of the system, $c_0$.  The default parameters in Tables~\ref{tab:DefParam} are used with $a=\delta_0/100$, $\eta={4}/{9}$, and $\phi$ is
averaged over 16 values.  The number of center bead oscillations varies
with $\omega$, but is chosen such that the associated velocity converges.
$N$ and $t_f$ also vary with $\omega$ and are chosen such that no
perturbation reaches the boundaries. The different regime classes are
distinguished as indicated in the text.}
\label{fig:Freq-Vel}
\end{figure}

For $\omega>\omega_c$, neither of the plane waves comprising $u_{i^*}(t)$ can propagate.  As a result, the spatial and temporal DRT behavior breaks down.  This can be discerned in Figs.~\ref{fig:Freq-DRT}, and \ref{fig:Freq-Vel} where the DRT magnitude and velocity (that we will now proceed to define more precisely)
approach zero as $\omega$ approaches $\omega_c$.


To explore the effects of parameter variation on the magnitude
of the DRT, we now establish an additional ratcheting metric.  For each of
the DRT profiles under consideration, $D_M$ represents the maximal
amount of temporal ratcheting at any given time.  We use this value as the
temporal DRT metric.  On the other hand, when
spatial ratcheting is present, it is first observed by comparing beads adjacent to the $i^*$th bead, that is at $j=1$.   As time progresses, the behavior spreads out from the center bead and is also observed for larger values of $j$.  We observe that $D_j$ is approximately the same for each $j$  exhibiting spatial DRT behavior; therefore the magnitude of spatial ratcheting can be quantified by
means of $D_1$.  Both $D_M$ and $D_1$ vary slightly over time.
Nevertheless, we have verified that this variation is small and, therefore,
choose to measure $D_1$ and $D_M$ at the final integration time $t_f$.

Figure~\ref{fig:Freq-DRT} illustrates the magnitude of spatial
and temporal DRT, based on the above diagnostic,
for a wide range of representative frequencies.  
For Class I ($10<\omega<15$), the
magnitude of temporal DRT slowly increases with the frequency.  
This regime corresponds to both input frequencies of the
forcing ($\omega$ and $2\omega$) being below the cutoff
frequency $\omega_c = 40.31$.
As it can be noticed in Fig.~\ref{fig:Freq-DRT}, the spatial
DRT starts appearing once the second harmonic ($2\omega$) 
of the driver
gets closer to the cutoff frequency
(see left vertical dashed line). This seems to be an
effect of the nonlinear response of the system that 
``widens'' the region of the cutoff frequency.
In fact, Class II corresponds to the region of frequencies
where the second harmonic of the driver transitions from
being transmitted to completely being stopped due to the
cutoff frequency.
It is interesting that for
Class II and IIIA, defined by $15<\omega<32.5$ corresponding 
to $\omega<\omega_c$ but $2\omega$ close to (under or over) 
$\omega_c$, the temporal DRT magnitude is approximately constant.  
However, as the first harmonic ($\omega$) starts getting
close to the cutoff frequency (see right vertical dashed line), 
naturally, both spatial and
temporal DRT start to disappear and eventually vanish,
as expected, once both, first and second, driver harmonics are inside the 
forbidden gap.
The effect of the first harmonic starting to approach
the cutoff frequency begins at, approximately, $\omega = 32.5$, 
corresponding to the onset of
Class IIIB behavior, the magnitude of temporal DRT begins to sharply
decrease as $\omega$ approaches $\omega_c$.  This lack of DRT for higher
frequencies is consistent with the DRT breaking down for $\omega>\omega_c$.
On the other hand, spatial DRT significantly increases as we move
from Class I to Class II and subsequently IIIA, and it also, in turn,
sharply decreases in the case of Class IIIB.

We define the \emph{outer-cone horizon} as the location of the onset of temporal DRT and the \emph{inner-cone horizon} as the location of the onset of spatial DRT.  
The velocities of the outer and inner horizon
are calculated numerically for values of $\omega$ ranging from 10 to 40.
For each frequency, the velocity of the outer-cone horizon approximately 
corresponds to the sound velocity of the system, 
$c_0^2 \equiv 6A\delta_0^\frac{1}{2}r^2/m$ (see filled dots in Fig.~\ref{fig:Freq-Vel}).  
This is a consequence of the nonlinearity of the system that 
``mixes'' the frequencies introduced by the forcing and
thus excites all modes.
In contrast to the behavior observed by the
outer-cone velocity, as shown in Fig.~\ref{fig:Freq-Vel} (see open dots), 
the inner-cone
velocity strongly depends on the forcing frequency.  For small values of $\omega$, where both $\omega$ and $2\omega$ are smaller than $\omega_c$, there is essentially no spatial DRT, as the two cones propagate with essentially the same
speed forming the ``pulse'' observed in the asymmetry indicator $D_j$.
As the frequency increases through Class I and
toward Class II, the velocity of the inner-cone horizon decreases significantly until the threshold $2\omega>\omega_c$ is crossed.
At that point the inner-cone velocity approaches zero. Within Class II,
the continuously decreasing velocity of the inner-cone forms the
tail of the kink discussed in connection to Fig.~\ref{fig:ImpProf}.
Past the point of $\omega=\omega_c/2$,
 the inner-cone velocity abruptly increases to a value similar to that for the
lower frequencies.  Subsequently, in Class IIIA, the velocity
decreases as the case where both $\omega$ and $2\omega$ are larger than $\omega_c$ is approached.  Class IIIB corresponds to a larger rate of decreasing
velocity.  Interestingly, the spatio-temporal wave velocities are independent of the choice of $\eta$.

It is interesting to point out at this stage that the presence of DRT in our
system is a direct consequence of the symmetry breaking provided
by the external forcing when $\eta\not=0$ [see Eq.~(\ref{Biharmonic2})].
However, it is also important to note that nonlinearity is also
a key ingredient for the presence of DRT. 
In fact, if the Hertzian
forces in Eq.~(\ref{EOMComp}) are replaced by linear (Hooke) springs, DRT
is no longer present. The absence of DRT for the linear force case
is a consequence of the fact that the DRT corresponding
to a phase $\phi$ is the negative of the one for phase
$\phi+T/2$, where $T$ is the period of the driver, and thus
providing a cancellation when DRT is averaged through
all possible phases of the driver.
Finally, it is also important to stress that,
as discussed earlier and depicted in Fig.~\ref{fig:VaryAmp}, 
the magnitude of DRT is, approximately, proportional to the cube of the 
forcing amplitude. Therefore, nonlinearity in the system is not only
necessary for observing DRT, but it also provides a nonlinear enhancement
of the DRT magnitude with respect to the input amplitude.

\begin{figure}[htb]
\vspace{-0.2cm}
\centering
$\begin{array}{c}
\includegraphics[width = 5.5cm]{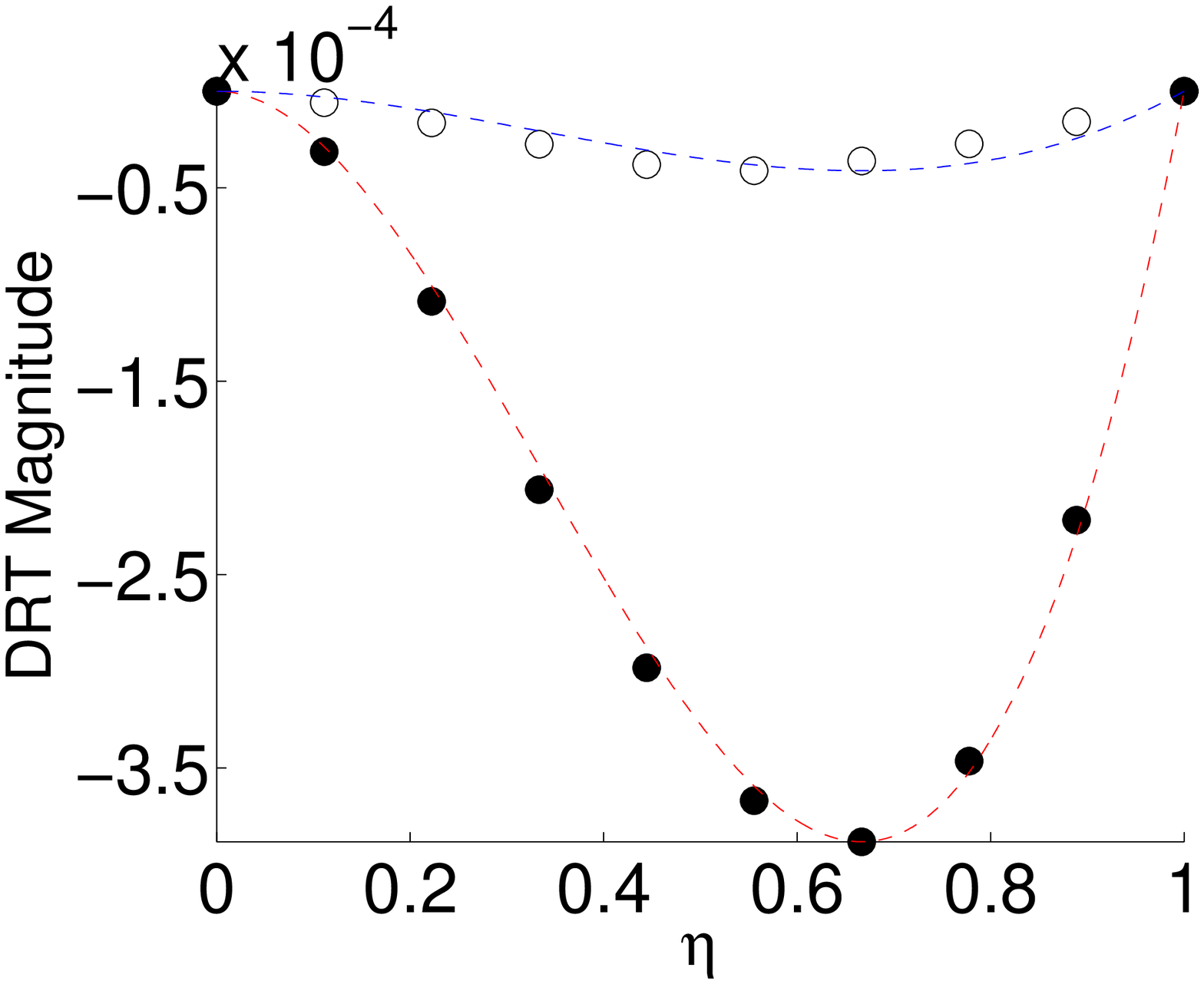} \\
\includegraphics[width = 5.5cm]{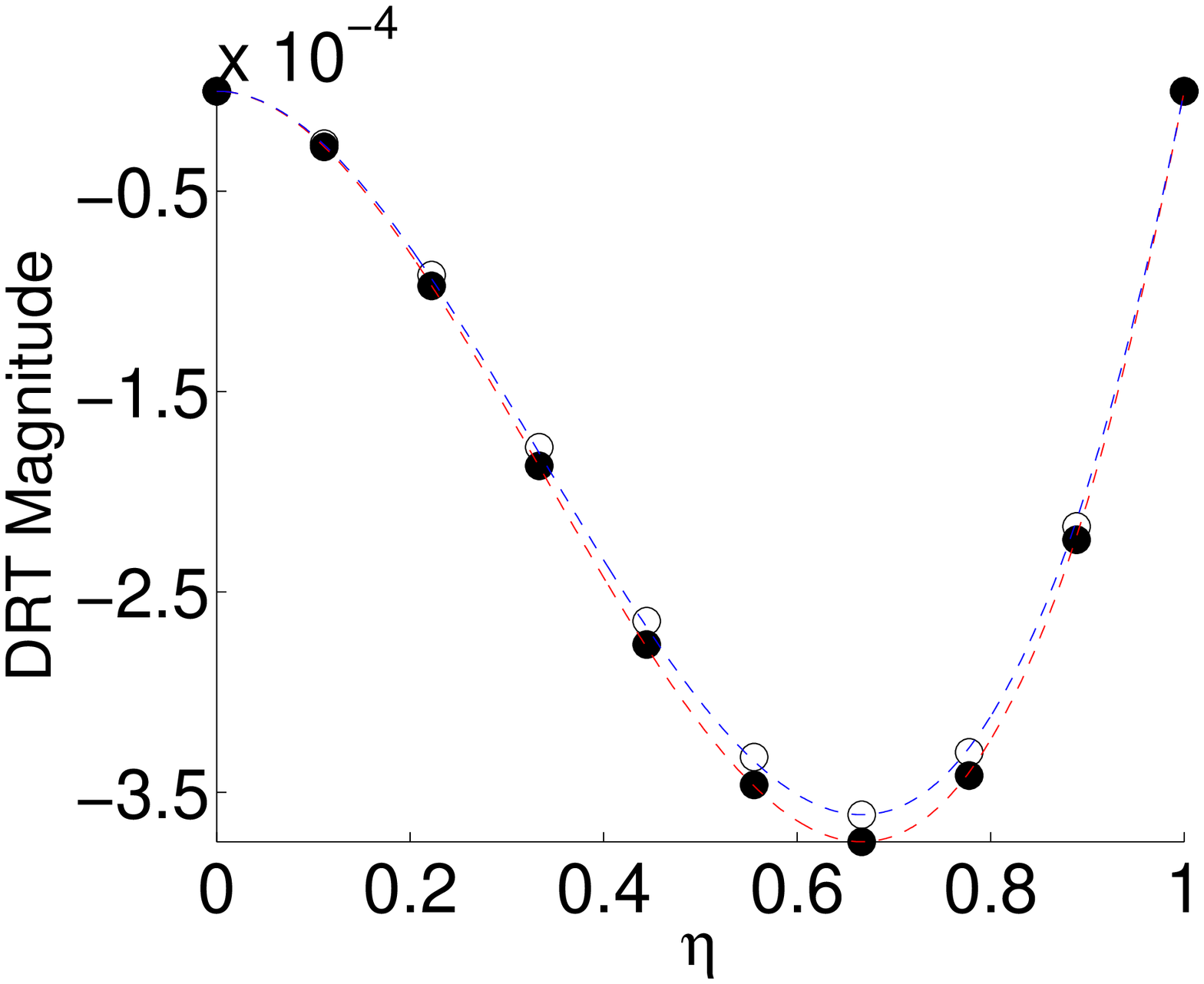}
\end{array}$
\caption{Magnitude of spatial and temporal DRT as a function of $\eta$. In the top panel, $\omega = 17.5$ while in the bottom $\omega = 30$.   The open dots represent the magnitude of spatial DRT, the filled dots are the magnitude of temporal DRT, and the dashed lines correspond to the theoretical ratchet velocity function.  The same parameter values and conditions as in Fig.~\ref{fig:Freq-Vel} were used.}
\label{fig:Eta-DRT}
\end{figure}

In Ref.~\cite{Chacon2007}, it was shown that when considering the forcing function $u_{i^*}(t)$, the magnitude of DRT behavior is due to two competing effects: the increase in the degree of symmetry breaking and the decrease in the transmitted impulse over a half-period, which is denoted as effective symmetry breaking.  It was shown that the DRT behavior is optimally enhanced for $\eta={2}/{3}$.  We now demonstrate that this result holds for the granular crystal case under consideration. We consider forcing frequencies of  $\omega$ = 17.5 and 30.
Figure~\ref{fig:Eta-DRT} illustrates the spatio-temporal DRT magnitude as a function of $\eta$.  As is expected, for $\eta=0$, the single harmonic does not induce DRT behavior.  As $\eta$ increases, the magnitude of DRT increases
until a maximum value for ratcheting is reached at $\eta = {2}/{3}$
(in our case since DRT is towards the left, the maximum ratcheting effect
corresponds to a minimum for DRT), 
with the exception of spatial DRT for $\omega$ = 17.5. The magnitude then decreases until DRT is again not present at $\eta=1$.  To further explore and quantitatively appreciate
this result, consider a generic biharmonic forcing function $f(t)=\epsilon_1\sin(q\omega t+\phi_1)+\epsilon_2\sin(p\omega t+\phi_2)$, where $\epsilon_{1,2}$ are amplitudes, $\phi_{1,2}$ are phases, $\omega$ is the frequency and $p$ and $q$ are coprimes.  It can be shown that the ratchet velocity $\bar{v}=\beta\left(\epsilon_1^p\epsilon_2^q\right)$ where $\beta$ is a system-dependent constant~\cite{Quintero2010}.  With the parameters in $u_{i^*}(t)$, we have $\bar{v}(\eta)=\beta\left(\eta^2-\eta^3\right)$.  Observe that this function has a minimum (for $\beta<0$) at $\eta={2}/{3}$, which is consistent with the numerical simulations.  By setting $\bar{v}({2}/{3})$ equal to the numerically-calculated DRT magnitude at $\eta={2}/{3}$, we can solve for the free parameter $\beta$.  These curves are shown in  Fig.~\ref{fig:Eta-DRT}.  There is a striking agreement between the calculated numerical DRT magnitudes and theoretical curves.  We see less of a correspondence for spatial DRT for $\omega=17.5$.  In general, for other frequencies considered, the temporal DRT matched the theoretical curves more consistently than spatial DRT, particularly for $\omega$ near $\omega_c/2$ and $\omega_c$.   

As a side note, it is possible to draw physical intuition for
the optimal biharmonic weight being at 2/3 if one considers 
the ideal ratcheting forcing: a sawtooth function.
Expanding a sawtooth function in Fourier series and keeping
only the first two harmonics it is straightforward to show that
their ratio is 2. In our case, when
$\eta=2/3$, we precisely get a ratio between the two harmonics
of $\eta/(1-\eta)=2$. In other words, the optimal ratcheting
forcing, i.e.~$\eta=2/3$, is the best possible approximation 
to a sawtooth function when using two harmonics.

\section{The Dissipative Bead-lattice}

\begin{figure}[htb]
\vspace{-0.2cm}
\includegraphics[width = 4.2cm]{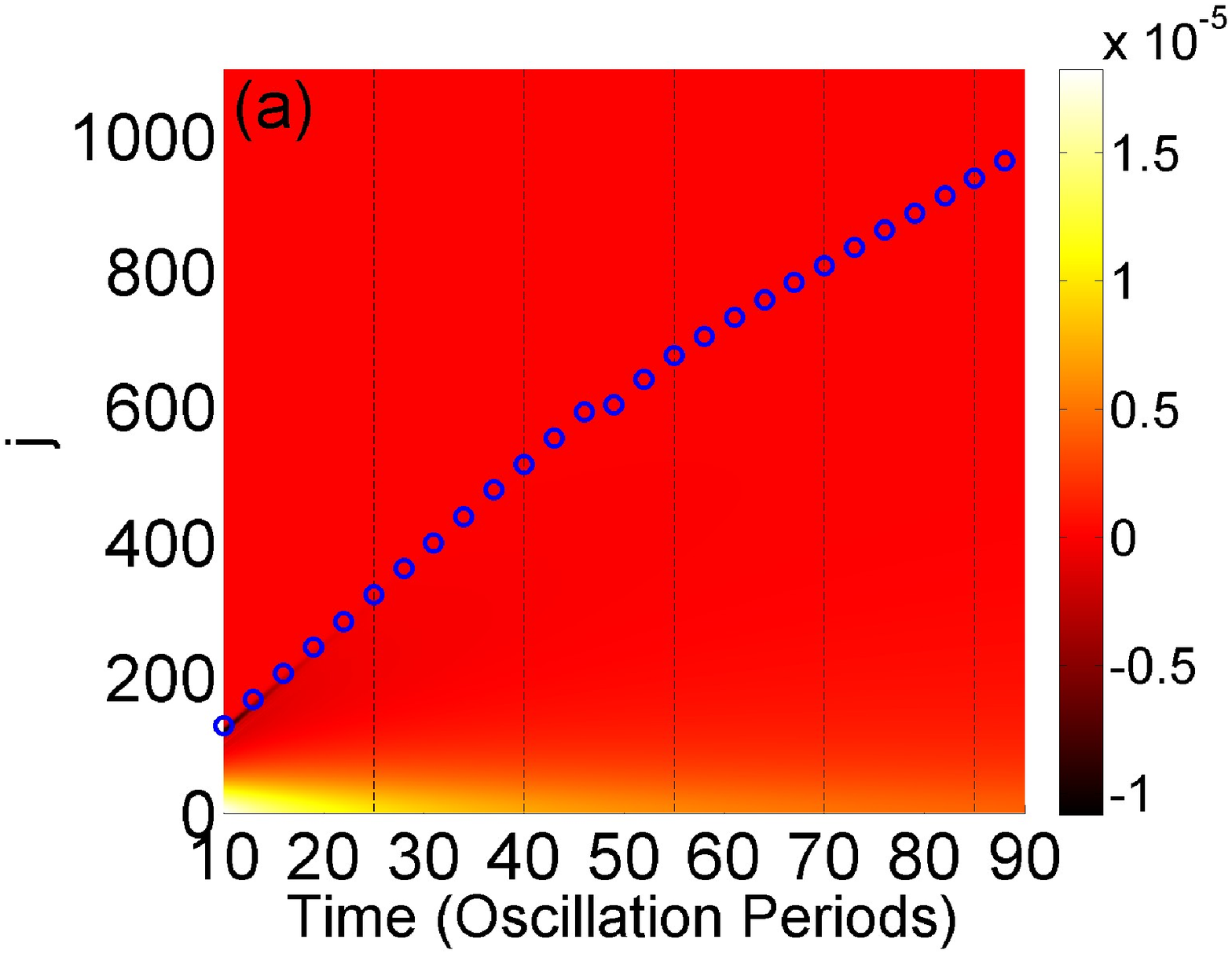}
\includegraphics[width = 4.2cm]{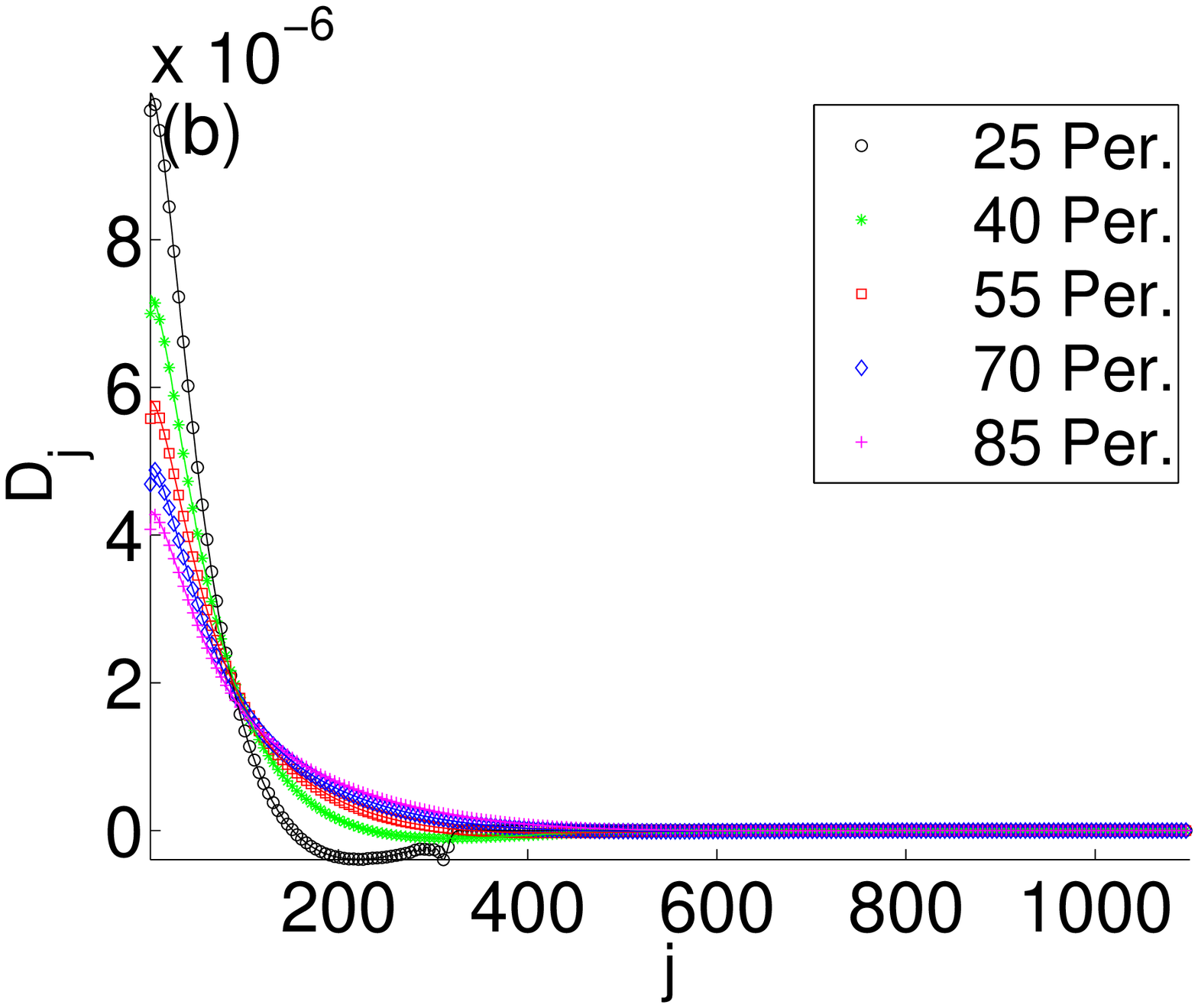}\\
\includegraphics[width = 4.2cm]{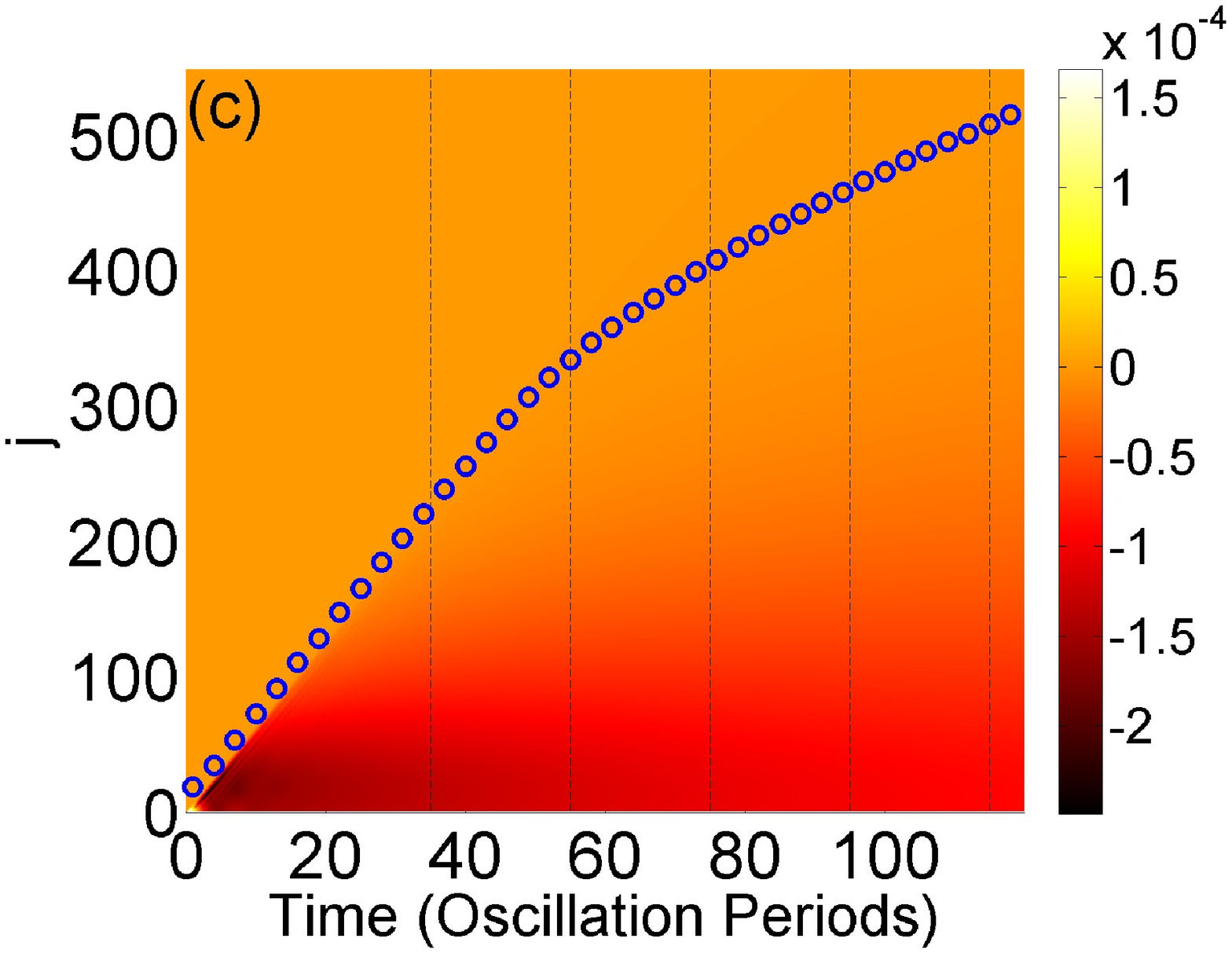}
\includegraphics[width = 4.2cm]{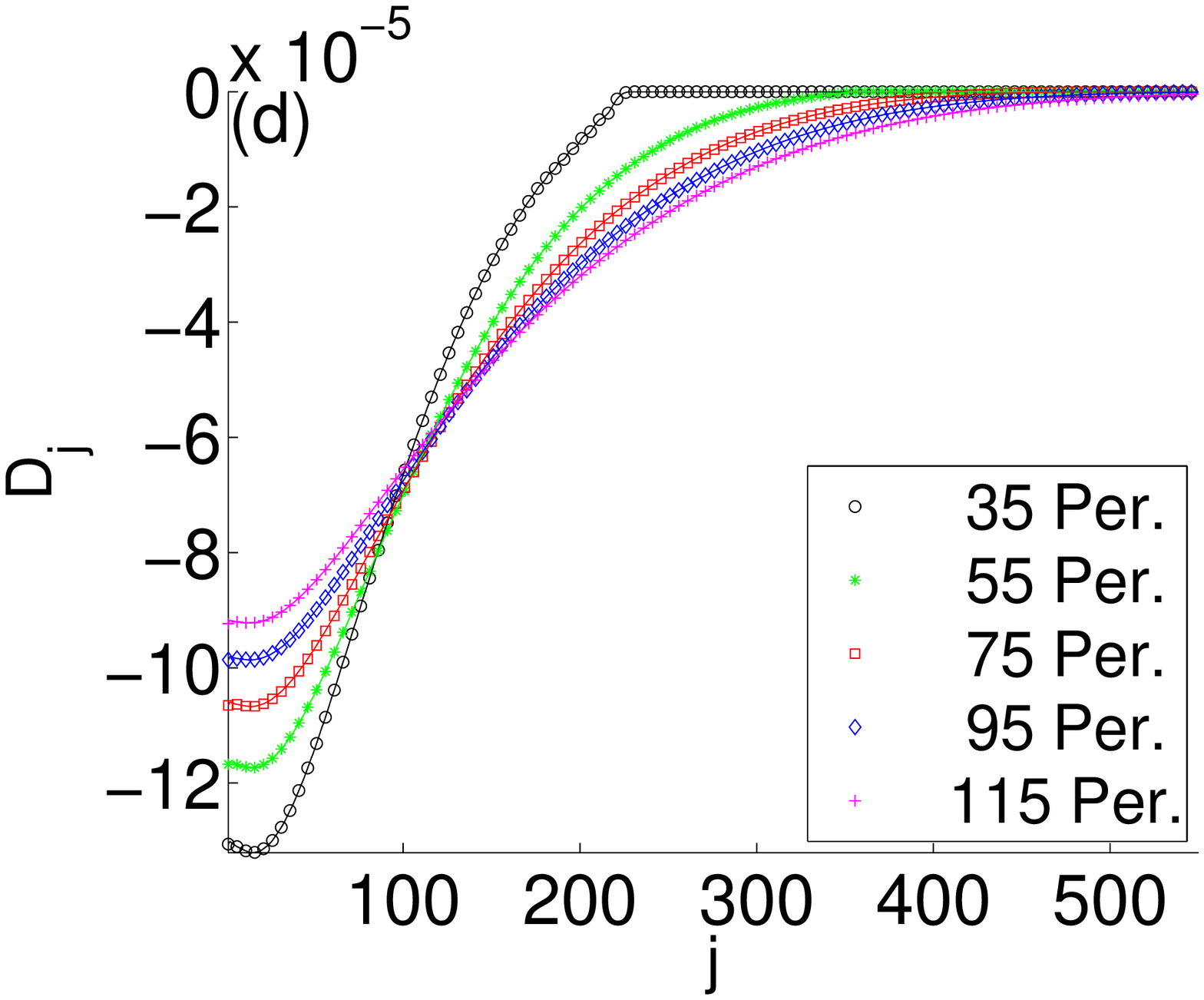}\\
\includegraphics[width = 4.2cm]{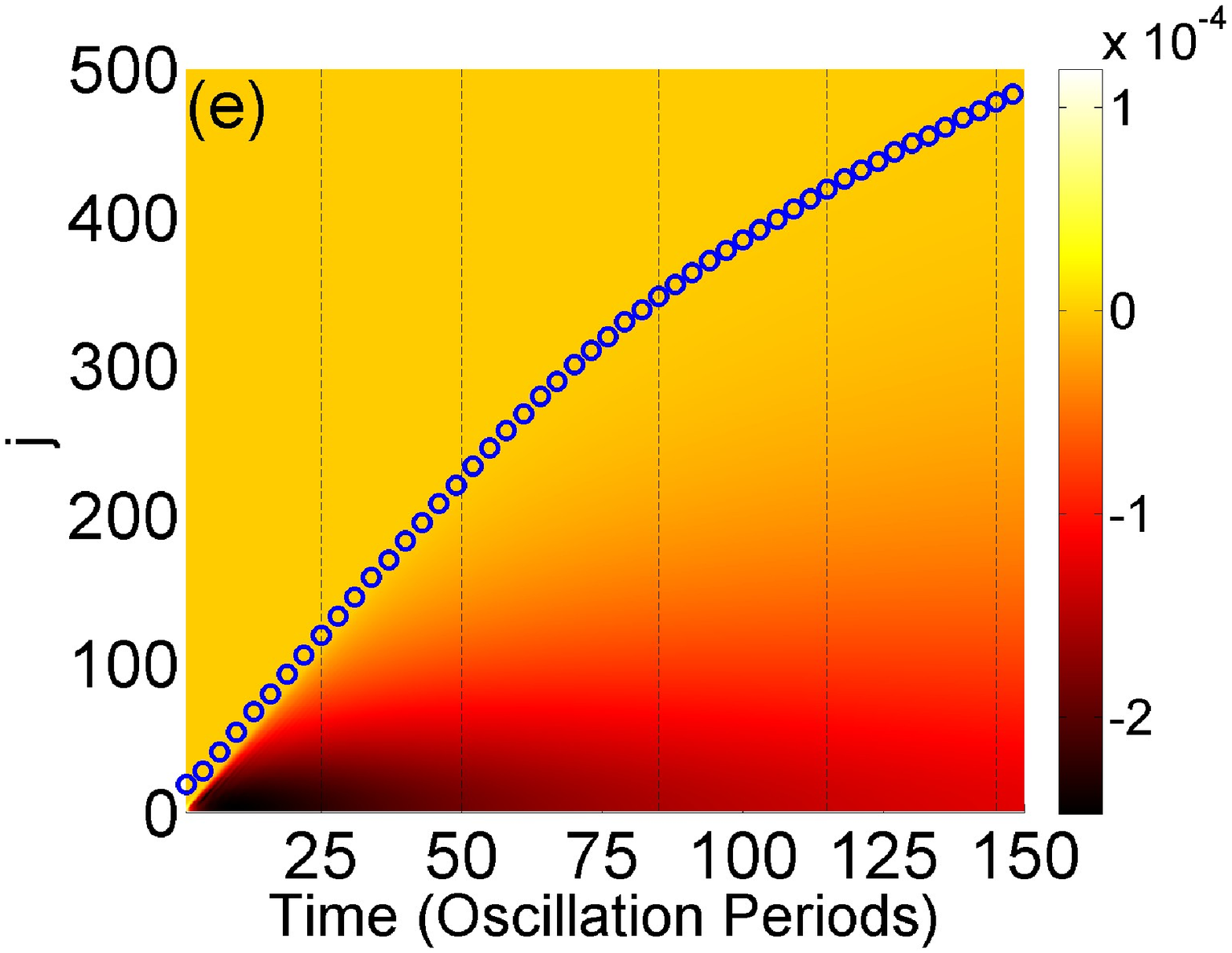}
\includegraphics[width = 4.2cm]{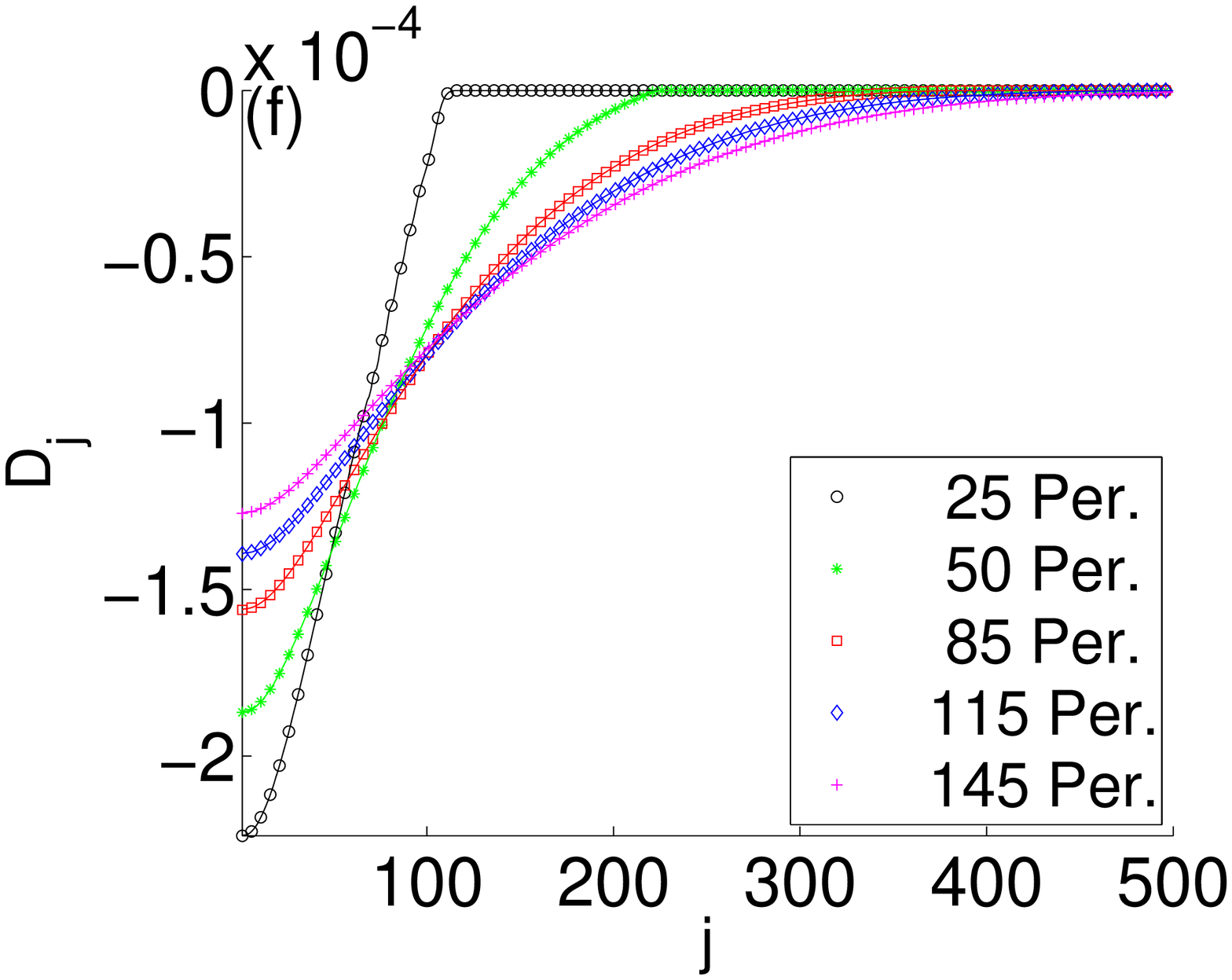}\\
\includegraphics[width = 4.2cm]{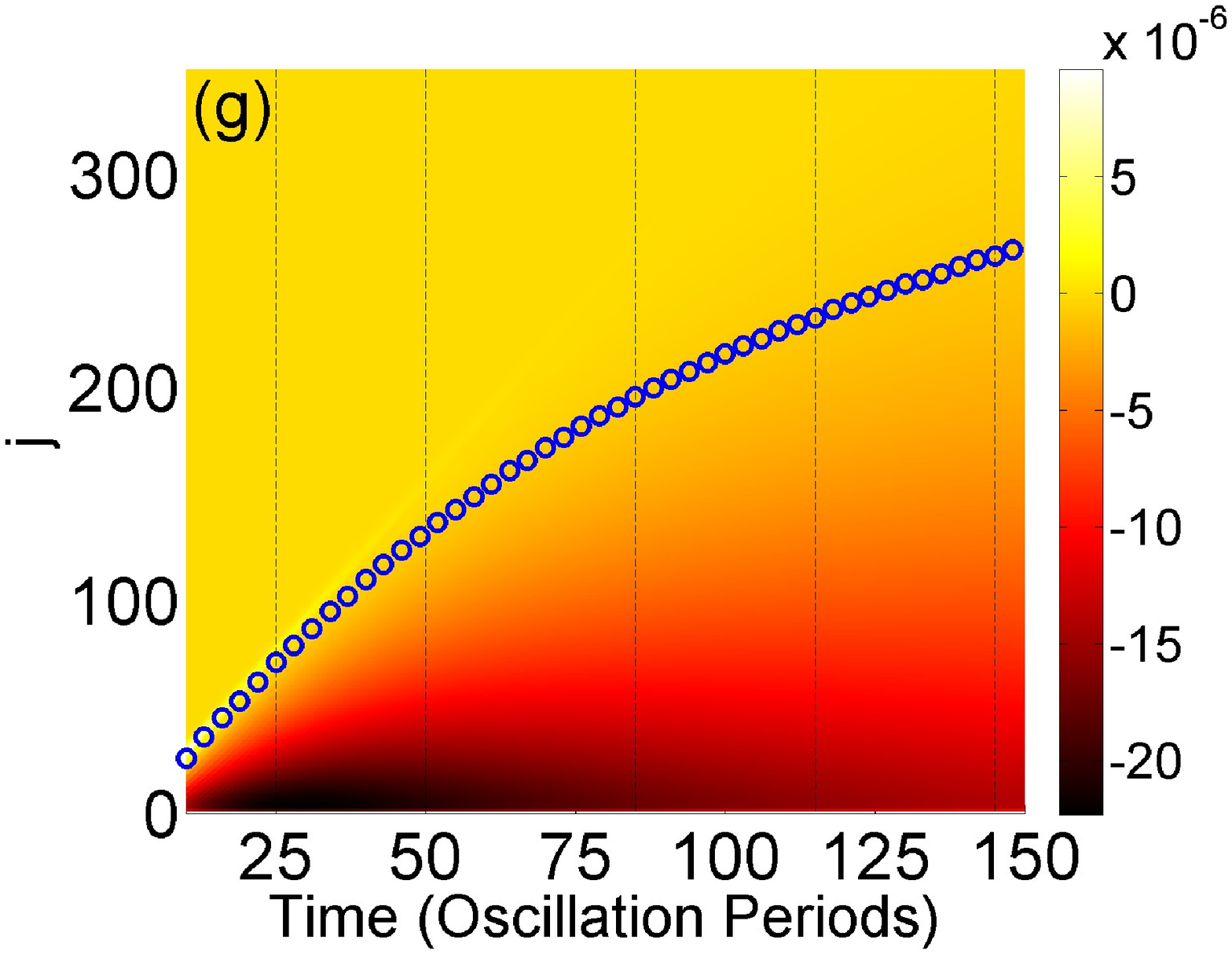}
\includegraphics[width = 4.2cm]{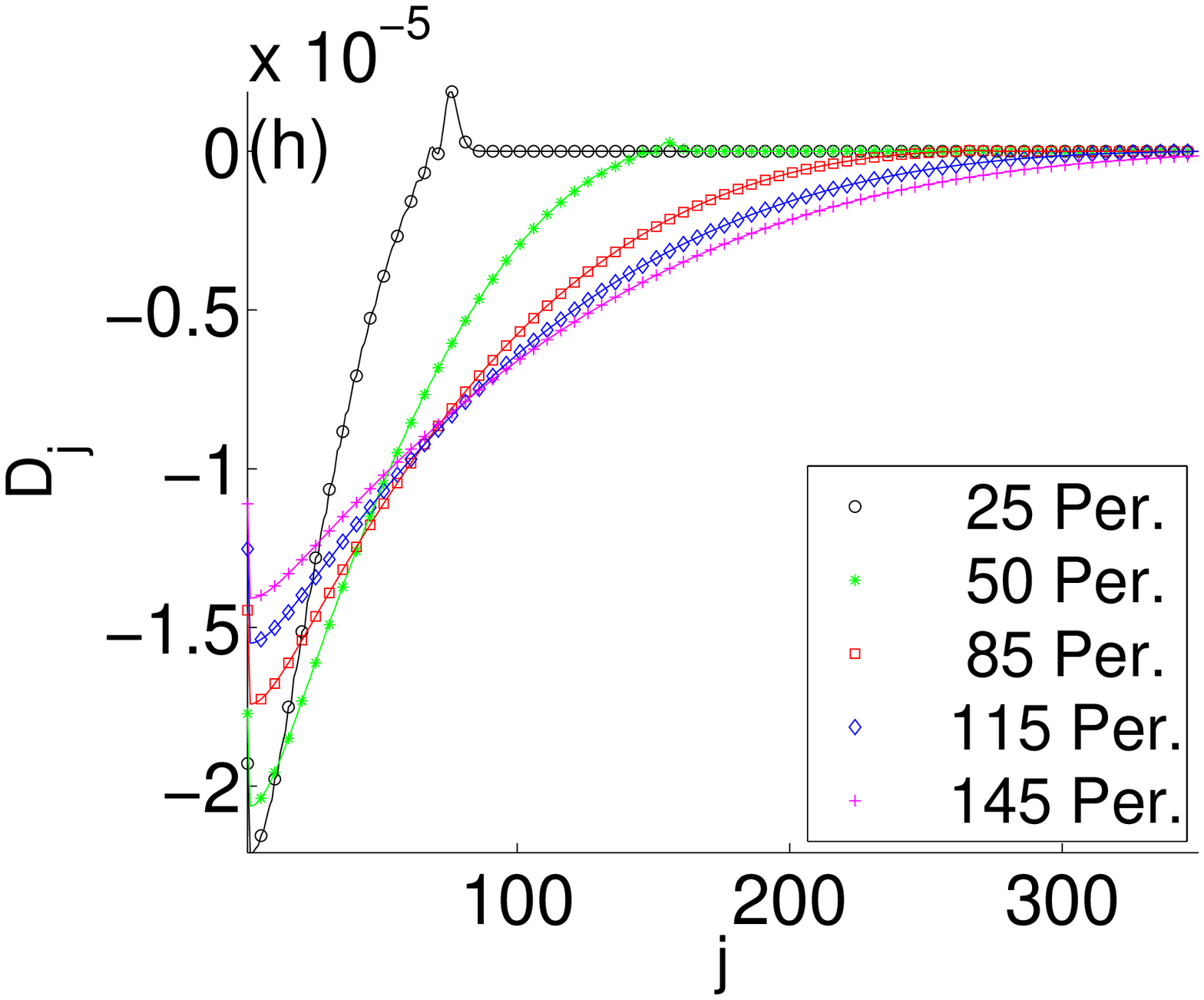}
\caption{Asymmetry indicator contour plots and temporal cross sections
in the presence of dissipation with $\tau=1750\mu s$.  The layout, parameters, and conditions are identical to Fig.~\ref{fig:ImpProf}.}
\label{fig:ImpProfFric}
\end{figure}

To more closely represent physical reality, dissipation is 
introduced into the uniform granular crystal by
augmenting Eq.~\eqref{EOMComp}:
\begin{equation}
m\ddot{u}_i = A[\delta_0+u_{i-1}-u_i]_+^\frac{3}{2} -
A[\delta_0+u_{i}-u_{i+1}]_+^\frac{3}{2}-\frac{m}{\tau}\dot{u}_i,
\end{equation}
where $\tau$ is a dissipation constant set equal to $1750\mu s$.  This value was chosen to match the dissipation constant used in the experiments of Ref.~\cite{Boechler2011}.  To investigate the effects of friction, we numerically integrate the four representative frequency cases presented earlier and illustrate
the corresponding results in  Fig.~\ref{fig:ImpProfFric}.
In order to allow sufficient time for transient behavior,
$t_f$, the final integration time, and $N$ were considerably larger than
for the non-friction simulations.

When friction is present, we observe that all cases
exhibit qualitatively similar behavior. Namely, unlike the 
frictionless case, the velocity of the temporal DRT is not 
set by the sound velocity of the system, but instead 
decreases as time progresses.  
Furthermore, in all dissipative cases, we have observed
that spatial DRT is weakened by the presence of
dissipation and that all cases present a similar
spatial structure which seems to involve a progressively widening
(i.e., dispersing) kink state.

It is worth noticing that for
$\omega=20$, the DRT profile is slightly different than 
for the other cases.  The maximal temporal DRT value, $D_M$ 
now occurs at $M\approx20$.  Furthermore, as time increases, 
$D_j\rightarrow D_M$ for $j<M$.  This is indicative of a small 
amount of spatial DRT.  In fact, the behavior is similar to 
Class IIIB for the frictionless system.
For $\omega=10$, the presence of friction radically changes the characteristics of the DRT profiles.  Temporal DRT is present, but for $t\gtrsim30$ periods, all non-zero values of $D_j$ are now positive, indicating that the DRT direction has \emph{reversed} and now favors left propagation.  For  $t\lesssim30$, the minimum of $D_j$ is less than zero illustrating the remnants of rightward DRT.

In each of the DRT profiles provided in Fig.~\ref{fig:ImpProfFric}, 
we observe two regimes of DRT propagation. Initially, the temporal 
DRT ``wave'' travels at the sound velocity of the system.  
However, as time progresses, the dissipation tend to slow
down the propagation of this wave.  
%
%
Similarly, dissipation is also responsible for progressively
damping out the DRT magnitude. Eventually, a steady state solution
is reached where the energy being pumped into
the system by the forcing is balanced by dissipation.

Finally, it is worth mentioning that, despite taking the optimal value
for the forcing amplitude to exploit maximum DRT gain, the DRT magnitude
is on the order of $10^{-5}$--$10^{-4}$ which amounts to a 
$10^{-3}$--$10^{-2}\%$ of biased transport between left and
right propagation.
Although these values are relatively small, DRT should be possible to 
measure in current experimental setups.

\section{Conclusions}
In this work, a one-dimensional granular chain (crystal) was considered
where the position of the center bead was prescribed by a biharmonic forcing function.   
This functional form is known to induce ratcheting.
Yet, in our case, a distinguishing characteristic was the
system-wide emergence  (i.e.~in space-time) of directed ratchet transport (DRT)
in the force profiles. The regimes where temporal (transient) ratcheting
and spatial (i.e., with a permanent spatial ``imprint'' over the lattice)
ratcheting were identified as a function of the system's frequency.
The relationship between the frequencies $\omega$ and $2\omega$ of the 
forcing function and the cutoff frequency $\omega_c$ of the system determined 
the characteristics of the observed DRT and its separation into different classes.
In the class I pertaining to temporal ratcheting,   a DRT ``wave" traveled
away from the center bead at the sound velocity of the system.  
Once the temporal DRT wave moved through a region, a
steady-state was induced in this region, wherein all bead pairs exhibited similar 
DRT magnitude.  If this value was non-zero, it corresponded to Class II and
the so-called spatial DRT.  The modification of the form of spatial ratcheting past
the regime where $\omega=\omega_c/2$ gave rise to yet another regime
that was referred to as Class III.

The frequency, $\omega$, and biharmonic weight, $\eta$ of the forcing function were varied so that the response of the magnitude of DRT and the velocity of the DRT ``waves" could be determined.  While the wave velocity was independent of the biharmonic weight, $\eta={2}/{3}$ maximized the magnitude of spatio-temporal DRT, in accordance with the expectations of Refs.~\cite{Chacon2007,Quintero2010}.

Friction was subsequently introduced into the system, leading to
weakening of the ratcheting effect and a rather
uniform spatial form of its profile in classes II and III.
Yet, it was class I that was most significantly affected by the inclusion of
friction within the system, which resulted in DRT switching directionality
from right to left.

Having paved the way for the consideration of ratchet effects in granular
crystals, there are numerous directions along which the present study
can be extended.
It is certainly of interest to attempt to expand the range of considered
materials and parameters (as well as that of heterogeneous systems
such as dimers, trimers~\cite{mason1,mason2,vakakis}) and of a wider range
of forcing frequencies and displacement parameters.
A key aspect of such a broader parametric effort is to try to maximize
the relevant DRT, so as to render it more accessible to potential experiments.
Another important direction of particular interest is to attempt to
expand the present considerations to the realm of higher dimensional
granular crystals. Recent efforts have made these gradually more
accessible to experimental investigations~\cite{andrea1,andrea2}
and hence such ratcheting efforts would be extremely timely and
relevant to consider. 


\acknowledgments{
Support from:
NSF DMS-0806762,
the US Air Force under grant FA9550-12-1-0332, Alexander S. Onassis,
and Alexander von Humboldt Foundation (P.G.K.),
NSF DMS-0806762 (R.C.G.),
NSF CMMI-1000337
and
NSF CAREER CMMI-844540 (C.D.\ and J.L.)
is kindly acknowledged.
}

\end{document}